\documentclass[12pt]{article}
\usepackage{amsmath,amssymb}
\usepackage{bm}
\usepackage{graphicx,color}
\usepackage{wrapfig}
\begin{document}
\title{
\begin{flushright}
\ \\*[-80pt] 
\begin{minipage}{0.2\linewidth}
\normalsize
%arXiv:YYMM.NNNN \\
\end{minipage}
\end{flushright}
{\Large \bf    $K_L \to \pi^0 \nu {\bar \nu}$ decay correlating with  $\epsilon_K$\\
 in  high-scale SUSY
\\*[20pt]}}
\author{
\centerline{Morimitsu~Tanimoto\footnote{E-mail address: tanimoto@muse.sc.niigata-u.ac.jp} \ \ and \ \
Kei~Yamamoto\footnote{E-mail address: yamamoto@muse.sc.niigata-u.ac.jp}}
\\*[20pt]
\centerline{
\begin{minipage}{\linewidth}
\begin{center}
{\it \normalsize
Department of Physics, Niigata University,~Niigata 950-2181, Japan }
\end{center}
\end{minipage}}
\\*[70pt]}

\date{
\centerline{\small \bf Abstract}
\begin{minipage}{0.9\linewidth}
\vskip  1 cm
\small
We have studied the contribution of the high-scale SUSY to the 
 $K_L \to \pi^0 \nu{\bar \nu}$ and $K^+ \to \pi^+ \nu{\bar \nu}$ processes
by  correlating with the CP violating parameter $\epsilon_K$.
Taking account of the recent LHC results for the Higgs discovery and the SUSY searches,
 we consider the high-scale SUSY at the $10-50$TeV scale
 in the framework of  the non-minimal squark (slepton) flavor mixing.
The Z penguin mediated the chargino dominates the SUSY contribution
 for these decays.
 At the $10$TeV scale of the  SUSY, the chargino contribution can enhance the branching ratio
of $K_L \to \pi^0 \nu {\bar \nu}$
in eight times compared with the SM predictions whereas 
  the predicted branching ratio $BR(K^+ \to \pi^+ \nu {\bar \nu})$
increases up to three times of the SM one.
The gluino box diagram
dominates the SUSY contribution of   $\epsilon_K$ up to $30\%$.
If the down-squark mixing is neglected compared with  the up-squark mixing,
the Z penguin mediated the chargino  dominates both
  SUSY contributions of   $BR(K_L \to \pi^0 \nu  {\bar \nu})$ and    $\epsilon_K$.
Then, it is found  a correlation between them, but
 the chargino contribution  to  $\epsilon_K$ is at most $3\%$.
 Even if the SUSY scale is  $50$TeV, 
 the chargino process still enhances the branching ratio of
$K_L \to \pi^0 \nu  {\bar \nu}$ from the SM prediction in the factor two, and 
  $\epsilon_K$ is deviated  from the SM prediction in ${\cal O}(10\%)$.
We also discuss the chargino contribution to $K_L \to \pi^0 e^+e^-$ process. 
\end{minipage}
}

\begin{titlepage}
\maketitle
\thispagestyle{empty}
\end{titlepage}

%%%%%%%%%%%%%%%%%%%%%%%%%%%%%%%%%%%%%%%%%%%%%%%%%%%%%%%%%%%
%%%%%%%%%%%%%%%%%%%%%%%%%%%%%%%%%%%%%%%%%%%%%%%%%%%%%%%%%%%
\section{Introduction}
%%%%%%%%%%%%%%%%%%%%%%%%%%%%%%%%%%%%%%%%%%%%%%%%%%%%%%%%%%%
%%%%%%%%%%%%%%%%%%%%%%%%%%%%%%%%%%%%%%%%%%%%%%%%%%%%%%%%%%%
\label{sec:Intro}

The $K$ meson physics have provided  important informations in the  indirect search for New Physics (NP). Especially, the rare decay processes 
$K^+ \to \pi^+ \nu {\bar \nu}$ and $K_L \to \pi^0 \nu  {\bar \nu}$ are known as the clean one theoretically \cite{Buras:1998raa,Buras:2015qea}.
Therefore, these both processes  have been  considered to be one of the powerful probes of NP
\cite{Bertolini:1986bs}-\cite{Smith:2014mla}
 whereas these decay widths  are  bounded by  so called  the  Grossman-Nir bound 
 for the  NP \cite{Grossman:1997sk,Fuyuto:2014cya}.

The  $K_L \to \pi^0 \nu  {\bar \nu}$ process  is 
the CP violating one   
and provides the direct measurement of the CP violating phase 
in the Cabibbo-Kobayashi-Maskawa (CKM) matrix \cite{Cabibbo:1963yz,Kobayashi:1973fv}.
In addition, the CP conserving process $K^+ \to \pi^+ \nu {\bar \nu}$ is also the physical quantity related with the unitarity triangle (UT). 
On the other hand, the  CP violating parameter $\epsilon_K$, 
which is  induced by the $K^0-{\bar K^0}$ mixing, 
also constrains the height of the UT.   
Hence these measured variables   give us the information of the UT fit as well as the CP violating quantity $\sin 2 \phi_1$ induced by the $B^0 -{\bar B^0}$ mixing. 
Furthermore, the $K \to \pi \nu{\bar \nu}$ processes are expected  
to open the NP window in the CP violating flavor structure.

The $K^+ \to \pi^+ \nu {\bar \nu}$ and $K_L \to \pi^0 \nu  {\bar \nu}$ decay processes 
are governed by the $Z$ penguin diagram  in the Standard Model (SM) \cite{Brod:2010hi},
which predicts
\begin{align}
BR(K_L \to \pi^0 \nu  {\bar \nu})_{\rm SM}
&=(2.43_{-0.37}^{+0.40} \pm 0.06)\times 10^{-11} , \nonumber \\
BR(K^+ \to \pi^+ \nu {\bar \nu})_{\rm SM}
&=(7.81^{+ 0.80}_{-0.71} \pm 0.29)\times 10^{-11} \ .
\label{SMprediction}
\end{align}
In the estimation of the branching ratio of $K \to \pi \nu{\bar \nu}$, the hadronic matrix elements can be extracted with the isospin symmetry relation \cite{Marciano:1996wy,Mescia:2007kn}.  
These processes are theoretically clean because the long-distance contributions  are small
\cite{Buras:2004uu}, and then the theoretical uncertainty is estimated below several percent.
On the other hand,  $\epsilon_K$  has the different flavor mixing  structure from these processes
since it is induced by the box diagram of  $K^0-{\bar K^0}$ mixing.
Therefore, the NP  is expected to appear in both $K \to \pi \nu{\bar \nu}$ and $\epsilon_K$ 
with different magnitudes.

On the experimental side, the upper bound of the branching ratio of
 $K_L \to \pi^0 \nu  {\bar \nu}$ is given  by the KEK E391a experiment \cite{Ahn:2009gb}. 
The branching ratio of $K^+ \to \pi^+ \nu {\bar \nu}$ measured by the BNL E787 and E949 experiments is  consistent with the SM prediction
\cite{Artamonov:2009sz};
\begin{align}
BR(K_L \to \pi^0 \nu  {\bar \nu})_{\rm exp}
&< 2.6 \times 10^{-8} ~(90\% {\rm C.L.}), \nonumber\\
BR(K^+ \to \pi^+ \nu {\bar \nu})_{\rm exp}
&=(1.73^{+1.15}_{-1.05})\times 10^{-10}  . %(1.73 ^{+1.15}_{-1.05})\times 10^{-10} .
\end{align}
At present, the J-PARC KOTO experiment is an in-flight measurement of
 $K_L \to \pi^0 \nu  {\bar \nu}$ approaching to the SM predicted precision 
\cite{Togawa:2014gya,Shiomi:2014sfa}, 
while the CERN NA62 experiment \cite{VenelinKozhuharovfortheNA62:2014lca}
 studies the $K^+ \to \pi^+ \nu {\bar \nu}$ process.

On the theoretical side,  the supersymmetry (SUSY) is one of the most attractive candidates for the NP.  However, the SUSY signals have not been observed yet, and then
  the recent searches for new particles at the LHC give us important constraints for the SUSY.
Since the lower bounds of masses of the SUSY particles  increase gradually, 
the squark and the gluino masses are supposed  to be at the higher scale than $1$ TeV
 \cite{Aad:2014wea,Chatrchyan:2014lfa,Aad:2014kra}.
Moreover, the SUSY models have been seriously constrained  by the Higgs discovery, in which
the Higgs mass is  $126$ GeV~\cite{Aad:2012tfa,Chatrchyan:2012ufa}. 

These facts suggest a class of SUSY models with heavy sfermions.
If  the squark and slepton masses
are expected to be  ${\cal O}(10-100)$ TeV,
 the lightest Higgs mass can be pushed up to $126$ GeV, whereas all SUSY particles  can be out of the reach of the LHC experiment.
Therefore, the indirect search of the SUSY particles becomes important in the low energy
flavor physics \cite{Altmannshofer:2013lfa,Moroi:2013sfa,Tanimoto:2014eva}.

So far, the effects of SUSY on the $K^+ \to \pi^+ \nu {\bar \nu}$ and $K_L \to \pi^0 \nu  {\bar \nu}$ processes have been studied in the framework of the Minimal Supersymmetric Standard Model (MSSM)  with the minimal flavor violation (MFV) scenario intensively 
\cite{Goto:1998qv,Buras:2000qz}.
Since the SUSY mass scale is pushed up more than $1$ TeV region at present, 
the effect of the  MSSM with MFV is expected to be very small.
These processes  are also discusses in the framework of the general SUSY model
\cite{Buras:1999da,Nir:1997tf,Buras:1997ij,Colangelo:1998pm,Nir:2002ah,Buras:2004qb,Isidori:2006qy} at the ${\cal O}(500)$ GeV scale.

 We have studied the SUSY contribution to the CP violation of the $B$ meson 
 and $\epsilon_K$ induced by the $K^0-{\bar K^0}$ mixing 
 under the relevant SUSY particle spectrum constrained by the observed Higgs mass 
\cite{Tanimoto:2014eva}.
 Then, it is found that the SUSY contribution  could be up to  $40\%$
 in the observed  $\epsilon_K$, on the other hand,
 it is minor in the CP violation of the $B$ meson
 at the high scale of $10-50$ TeV. 
 Therefore, in this paper, we investigate the high-scale SUSY contribution 
to $K^+ \to \pi^+ \nu {\bar \nu}$ and $K_L \to \pi^0 \nu  {\bar \nu}$  by correlating with  $\epsilon_K$ in the framework of the mass eigenstate of the SUSY particles, 
 which is consistent with the  updated experimental situations like the direct SUSY searches and the Higgs discovery, with the non-minimal squark (slepton) flavor mixing.

Our paper is organized as follows.
Sec.2 gives the basic framework of $K^+ \to \pi^+ \nu {\bar \nu}$, $K_L \to \pi^0 \nu  {\bar \nu}$ and $\epsilon_K$ in the SM and the MSSM.
In Sec.3, we present  the setup  of the high-scale SUSY.
In Sec.4, we discuss  our numerical   results.
Sec.4 is devoted to the  summary. 
The SUSY mass spectra and the Z penguin amplitude mediated the chargino are given in Appendices A and B, respectively.

%%%%%%%%%%%%%%%%%%%%%%%%%%%%%%%%%%%%%%%%%%%%%%%%%%%%%%%%%%%
%%%%%%%%%%%%%%%%%%%%%%%%%%%%%%%%%%%%%%%%%%%%%%%%%%%%%%%%%%%
\section{Basic framework}
%%%%%%%%%%%%%%%%%%%%%%%%%%%%%%%%%%%%%%%%%%%%%%%%%%%%%%%%%%%
%%%%%%%%%%%%%%%%%%%%%%%%%%%%%%%%%%%%%%%%%%%%%%%%%%%%%%%%%%%
\label{sec:Pre}
In this section, we present the basic formulae for the  $K \to \pi \nu {\bar \nu}$ decay  and  the CP violating parameter $\epsilon_K$, which correspond to $|\Delta S|=1$ and $|\Delta S|=2$ processes, respectively.
The $K^+ \to \pi^+ \nu {\bar \nu}$ and $K_L \to \pi^0 \nu  {\bar \nu}$ processes are clean ones  theoretically since the hadronic matrix elements can be extracted including isospin breaking corrections by taking the ratio to the leading semileptonic decay of  $K^+ \to \pi^0 e^+ \nu$.
Moreover, the long-distance contributions to these rare decays are negligibly small.
Therefore, the accurate measurements of these  decay processes  provide the crucial tests of the  SM.  Especially, the $K_L \to \pi^0 \nu  {\bar \nu}$ process is purely the CP violating one, which can reveal the source of  the CP violating phase.

On the other hand, the CP violating parameter $\epsilon_K$ is  measured with enough accuracy.
The major theoretical ambiguity comes from the hadronic matrix element factor $\hat{B}_K$. 
The recent lattice calculations give us the reliable value for $\hat{B}_K$ 
\cite{Bae:2013lja,Aoki:2013ldr}.  
The more accurate estimate of the SM contribution  enables us to search the NP such a SUSY
because we know the accurate observed value of  $\epsilon_K$. 
Actually,  the non-negligible SUSY contribution has been  expected  in $\epsilon_K$ 
 at the scale of ${\mathcal O}(100)$ TeV \cite{Altmannshofer:2013lfa,Moroi:2013sfa,Tanimoto:2014eva}.
Consequently, it is required to examine the high-scale SUSY contribution in 
 $K \to \pi \nu {\bar \nu}$   by correlating with   $\epsilon_K$.

%%%%%%%%%%%%%%%%%%%%%%%%%%%%%%%%%%%%%%%%%%%%%%%%%%%%%%%%%%%
\subsection{Basic framework : $K^+ \to \pi^+ \nu {\bar \nu}$ and $K_L \to \pi^0 \nu  {\bar \nu}$ }
%%%%%%%%%%%%%%%%%%%%%%%%%%%%%%%%%%%%%%%%%%%%%%%%%%%%%%%%%%%

\subsubsection{$K^+ \to \pi^+ \nu {\bar \nu}$ and $K_L \to \pi^0 \nu  {\bar \nu}$ in the SM}
Let us start with discussing the framework of the 
$K^+ \to \pi^+ \nu {\bar \nu}$ and $K_L \to \pi^0 \nu  {\bar \nu}$ processes 
in the SM \cite{Buras:1998raa}.
The effective Hamiltonian for $K \to \pi \nu {\bar \nu}$ in the SM is given: 
\begin{align}
	{\mathcal H_{\rm eff}^{\rm SM}}
	=\frac{G_F}{\sqrt 2} \frac{2 \alpha}{\pi {\rm sin}^2\theta_W}
	\sum_{i=e, \mu, \tau} \left[ V_{cs}^*V_{cd} X_c + V_{ts}^*V_{td} X_t\right]
	\left( {\bar s_L}\gamma^{\mu} d_L \right)\left ({\bar \nu_L^i}\gamma_{\mu} \nu_L^i \right)+{\rm H.c.},
	\label{effH}
\end{align}
which is  induced by the box  and the Z penguin mediated the W boson. 
The dominant box contrition is derived by the top-quark exchange, on the other hand, 
 the charm-quark exchange  contributes to the Z penguin process as well as the top-quark one.
The up-quark contribution is negligible due to its small mass.
So, the loop function $X_c$ denotes the charm-quark contribution of the  Z penguin, and $X_t$ is the sum of the top-quark exchanges of the  box diagram and the  Z penguin in Eq.(\ref{effH}).

%%%%%%%%%%%%%%%%%%%%%%%%%%%%%%%%%%%%%%%%%%%%%%%%%%%%%%%
%%%%%%%%%%%%%%%%  K^+ -> pi^+ nu nu  %%%%%%%%%%%%%%%%%%
%%%%%%%%%%%%%%%%%%%%%%%%%%%%%%%%%%%%%%%%%%%%%%%%%%%%%%%
Let us define the function $F$ as follows: 
\begin{align}
	F=V_{cs}^*V_{cd} X_c+V_{ts}^*V_{td} X_t \ .
\end{align}
The branching ratio of $K^+ \to \pi^+ \nu {\bar \nu}$ is given in terms of $F$.
Taking 
the ratio of it to the branching ratio of  $K^+ \to \pi^0 e^+ {\bar \nu}$, which is the tree level process, we obtain a simple form:
\begin{align}
	\frac{BR(K^+ \to \pi^+ \nu {\bar \nu})}{BR(K^+ \to \pi^0 e^+ {\bar \nu})}
	=\frac{2}{|V_{us}|^2} 
	\left(\frac{\alpha}{2\pi {\rm sin}^2\theta_W}\right)^2
	\sum_{i=e, \mu, \tau} |F|^2 .
	\label{BrKp}
\end{align}
The  $K^+ \to \pi^0 e^+ {\bar \nu}$ decay is precisely measured as 
$BR(K^+ \to \pi^0 e^+ {\bar \nu})_{\rm exp} = (5.07 \pm 0.04)\times10^{-2}$ \cite{PDG}, and 
its hadronic matrix element is related to the one of $K^+ \to \pi^+ \nu {\bar \nu}$ with the isospin symmetry:
\begin{align}
	\langle \pi^0 | \left( {\bar d_L}\gamma^{\mu} s_L \right) | {\bar K^0}\rangle
	&=\langle \pi^0 | \left( {\bar s_L}\gamma^{\mu} u_L \right) | K^+ \rangle , \\
	\langle \pi^+ | \left( {\bar s_L}\gamma^{\mu} d_L \right) | K^+\rangle
	&={\sqrt 2}\langle \pi^0 | \left( {\bar s_L}\gamma^{\mu} u_L \right) | K^+ \rangle ,
\end{align}
where the coefficients are determined by the Clebsch-Gordan coefficient.
By using  this relation,  the hadronic matrix element has been  removed in Eq.(\ref{BrKp}).

Now the branching ratio for $K^+ \to \pi^+ \nu {\bar \nu}$ is expressed 
as follows:
\begin{align}
	BR(K^+ \to \pi^+ \nu {\bar \nu})
	=3 \kappa \cdot  r_{K^+} |F|^2,
	 % [({\rm Re}F)^2+({\rm Im}F)^2],
	\label{BRKpPipnunu}
\end{align}
\begin{align}
	\kappa
	=\frac{2}{|V_{us}|^2} 
	\left(\frac{\alpha}{2\pi {\rm sin}^2\theta_W}\right)^2 BR(K^+ \to \pi^0 e^+ {\bar \nu}),
\end{align}
where $r_{K^+}$ is the isospin breaking correction
 between $K^+ \to \pi^0 e^+ {\bar \nu}$ and $K^+ \to \pi^0 e^+ {\bar \nu}$ 
\cite{Marciano:1996wy,Mescia:2007kn}, and 
the factor 3 comes from the sum of three neutrino flavors.
It is noticed that the branching ratio for $K^+ \to \pi^+ \nu {\bar \nu}$ depends on both the real and imaginary part of  $F$.   
%%%%%%%%%%%%%%%%%%%%%%%%%%%%%%%%%%%%%%%%%%%%%%%%%%%%%%%
%%%%%%%%%%%%%%%%  K_L -> pi nu nu  %%%%%%%%%%%%%%%%%%%%
%%%%%%%%%%%%%%%%%%%%%%%%%%%%%%%%%%%%%%%%%%%%%%%%%%%%%%%

For the $K_L \to \pi^0 \nu  {\bar \nu}$ decay, 
the $K^0-{\bar K^0}$ mixing should be taken account, and one obtains
\begin{align}
	A(&K_L \to \pi^0 \nu  {\bar \nu})
	=\frac{G_F}{\sqrt 2} \frac{2 \alpha}{\pi {\rm sin}^2\theta_W}
	\left ({\bar \nu_L^i}\gamma_{\mu} \nu_L^i \right)
	\langle \pi^0| 
	\left[ F ({\bar s_L}\gamma_{\mu} d_L)+F^* ({\bar d_L}\gamma_{\mu} s_L) \right]
	 | K_L\rangle \nonumber \\ 
	&=\frac{G_F}{\sqrt 2} \frac{2 \alpha}{\pi {\rm sin}^2\theta_W}
	\left ({\bar \nu_L^i}\gamma_{\mu} \nu_L^i \right)
	\frac{1}{\sqrt 2}
	\left[ F(1+\bar{\epsilon})\langle \pi^0| ({\bar s_L}\gamma_{\mu} d_L) | K^0\rangle
	+F^*(1-\bar{\epsilon})\langle \pi^0| ({\bar d_L}\gamma_{\mu} s_L) | \bar K^0\rangle
	\right] \nonumber \\
	&=\frac{G_F}{\sqrt 2} \frac{2 \alpha}{\pi {\rm sin}^2\theta_W}
	\left ({\bar \nu_L^i}\gamma_{\mu} \nu_L^i \right)
	\frac{1}{\sqrt 2}
	\left[ F(1+\bar{\epsilon})-F^*(1-\bar{\epsilon})\right]
	\langle \pi^0| ({\bar d_L}\gamma_{\mu} s_L) | K^0\rangle \nonumber \\
	&\simeq \frac{G_F}{\sqrt 2} \frac{2 \alpha}{\pi {\rm sin}^2\theta_W}
	\left ({\bar \nu_L^i}\gamma_{\mu} \nu_L^i \right)
	\frac{1}{\sqrt 2} \
	2{\rm Im}F \
	\langle \pi^0| ({\bar d_L}\gamma_{\mu} s_L) | K^0\rangle .
\label{A(KLpi0nun)}
\end{align}
In the step of the first line going to the second line in (\ref{A(KLpi0nun)}) , we use
\begin{align}
	 | K_L\rangle = \frac{1}{\sqrt 2}
	\left[ (1+\bar{\epsilon}) | K^0\rangle
	+(1-\bar{\epsilon}) | K^0 \rangle \right] ,
\end{align}
and then, 
 after using  the CP transition relation in the second line,
\begin{align}
	{\rm CP} | K^0\rangle
	=-| \bar K^0\rangle, \ \ \ \ \
	{\rm C} | K^0\rangle
	=| \bar K^0\rangle ,
\end{align}
\begin{align}
	\langle \pi^0| ({\bar d_L}\gamma_{\mu} s_L) | \bar K^0\rangle
	=-\langle \pi^0| ({\bar s_L}\gamma_{\mu} d_L) | K^0\rangle ,
\end{align}
we obtain the equation in the  third line.
In the final line, we neglect the CP violation in $K^0-{\bar K^0}$ mixing, ${\bar \epsilon}$, due to its smallness $|{\bar \epsilon}| \sim 10^{-3}$.

Taking the ratio between the branching ratio of  $K^+ \to \pi^0 e^+ {\bar \nu}$
and  $K_L \to \pi^0 \nu  {\bar \nu}$, we have the simple form:
\begin{align}
	\frac{BR(K_L \to \pi^0 \nu  {\bar \nu})}{BR(K^+ \to \pi^0 e^+ {\bar \nu})}
	=\frac{2}{|V_{us}|^2} 
	\left(\frac{\alpha}{2\pi {\rm sin}^2\theta_W}\right)^2 \frac{\tau(K_L)}{\tau(K^+)} 
	\sum_{i=e, \mu, \tau} ({\rm Im} F)^2 .
\end{align}
Therefore,  the branching ratio of  $K_L \to \pi^0 \nu  {\bar \nu}$ is given as follows:
\begin{align}
	BR(K_L \to \pi^0 \nu  {\bar \nu})
	=3 \kappa \cdot \frac{r_{K_L}}{r_{K^+}}\frac{\tau (K_L)}{\tau (K^+)} ({\rm Im} F)^2 , 
\label{BRKLPi0nunu}
\end{align}
where $r_{K_L}$ and  $r_{K^+}$ denote the isospin breaking effect 
\cite{Marciano:1996wy,Mescia:2007kn}.
It is remarked that the branching ratio of $K_L \to \pi^0 \nu  {\bar \nu}$ depends on the imaginary part of $F$.   
Since the charm-quark contribution is negligible 
due to the small imaginary part of $V_{cs}^*V_{cd}$, it is enough to consider only the top-quark exchange in this decay.

In the SM, $K^+ \to \pi^+ \nu {\bar \nu}$ and $K_L \to \pi^0 \nu  {\bar \nu}$ are related
 to the UT fit.
We write down the branching ratio in terms of the Wolfenstein parameters.
Since ${\rm Re}F$ and ${\rm Im}F$ are given as 
\begin{align}
	{\rm Re}F
	= -\lambda X_c-A^2 \lambda^5 (1-\rho) X_t \ , \qquad
	{\rm Im}F
	= A^2 \lambda^5 \eta X_t \  ,
\end{align}
we can express  the branching ratio of these decays as
\begin{align}
	BR(K^+ \to \pi^+ \nu {\bar \nu})
	&=3\kappa \cdot  r_{K^+}
	[({\rm Re}F)^2+({\rm Im}F)^2] \nonumber \\
	&=3\kappa \cdot  r_{K^+} \cdot A^4 \lambda^{10} X_t^2 
	\Big[ \left( {\bar \rho}-\rho^0 \right)^2 +{\bar \eta}^2 \Big],
\label{ellipse}
\end{align}
where 
\begin{align}
\rho_0=1+\frac{X_c}{A^2 \lambda^4 X_t}  \ ,
\end{align}
and
\begin{align}
	BR(K_L \to \pi^0 \nu  {\bar \nu})
	&=3\kappa \cdot \frac{r_{K_L}}{r_{K^+}}\frac{\tau (K_L)}{\tau (K^+)} ({\rm Im} F)^2 \nonumber \\
	&=3\kappa \cdot \frac{r_{K_L}}{r_{K^+}}\frac{\tau (K_L)}{\tau (K^+)} \cdot A^4 \lambda^{10}  X(x_t)^2 \eta^2 .
\label{eta}
\end{align}
It is noticed that $BR(K^+ \to \pi^+ \nu {\bar \nu})$ in Eq(\ref{ellipse}) is approximately a 
%ellipse formula 
circle centered at ${\bar \rho} =\rho_0\simeq 1.2$, ${\bar \eta}=0$
 on the $\bar\rho$-$\bar\eta$ plane. 
On the other hand,  $BR(K_L \to \pi^0 \nu  {\bar \nu})$  in Eq(\ref{eta}) just depends on $\eta$ and it can determine the height of the UT directly.
In this way, the precise measurements of 
$K^+\to \pi^+\nu {\bar \nu}$ and $K_L \to \pi^0 \nu  {\bar \nu}$ become crucial tests for the SM.

Before going to discuss  the SUSY formulation, we present the general bound 
between $K^+ \to \pi^+ \nu {\bar \nu}$ and $K_L \to \pi^0 \nu  {\bar \nu}$, 
so called the Grossman-Nir bound \cite{Grossman:1997sk}.
As seen from above formulations, since the two processes are determined by the imaginary part and the absolute value of the same coupling, the  model independent bound is obtained:
\begin{align}
	BR(K_L \to \pi^0 \nu  {\bar \nu})
	< \frac{r_{K_L}}{r_{K^+}}\frac{\tau (K_L)}{\tau (K^+)} \cdot BR(K^+ \to \pi^+ \nu {\bar \nu}) 
	\lesssim 4.4 \times BR(K^+ \to \pi^+ \nu {\bar \nu})\ ,
\end{align}
where we use the isospin symmetry $A(K^+ \to \pi^+ \nu {\bar \nu})=\sqrt 2 A(K^0 \to \pi^0 \nu {\bar \nu})$.
This bound must be satisfied for any NP \cite{Grossman:1997sk,Fuyuto:2014cya}.

%%%%%%%%%%%%%%%%%%%%%%%%%%%%
%Finally, the SM predictions are given as \cite{Brod:2010hi}
%\begin{align}
%BR(K_L \to \pi^0 \nu  {\bar \nu})_{\rm SM}
%&=(2.43_{-0.37}^{+0.40} \pm 0.06)\times 10^{-11} , \\
%BR(K^+ \to \pi^+ \nu {\bar \nu})_{\rm SM}
%&=(7.81^{+ 0.80}_{-0.71} \pm 0.29)\times 10^{-11} \ ,
%\label{SMprediction}
%\end{align}
%on the other hand, the experimental data in the PDG \cite{PDG} are 
%\begin{align}
%BR(K_L \to \pi^0 \nu  {\bar \nu})_{\rm exp}
%&< 2.6 \times 10^{-8}, \nonumber\\
%BR(K^+ \to \pi^+ \nu {\bar \nu})_{\rm exp}
%&=(1.7 \pm 1.1)\times 10^{-10}  . %(1.73 ^{+1.15}_{-1.05})\times 10^{-10} .
%\end{align}
%%%%%%%%%%%%%%%%%%%%%%%%%%%%%%%%%%%

\subsubsection{$K^+ \to \pi^+ \nu {\bar \nu}$ and $K_L \to \pi^0 \nu  {\bar \nu}$ in  the MSSM}%%%%%%%%%%%%%

The effective Hamiltonian in Eq.(\ref{effH}) is modified due to   new box diagrams and penguin diagrams  induced by SUSY particles. Then, the effective Lagrangian  is given as 
\begin{align}
	{\mathcal L_{\rm eff}}
	=\sum_{i, j=e, \mu, \tau} \left[ C_{\rm VLL}^{ij} \left( {\bar s_L}\gamma^{\mu} d_L \right)
	+C_{\rm VRL}^{ij} \left({\bar s_R}\gamma^{\mu} d_R \right)\right]\left ({\bar \nu_L^i}\gamma_{\mu} \nu_L^j \right)+{\rm H.c.} \ ,
\end{align}
where $i$ and $j$ are the index of the flavor of the neutrino final state.
Here, $C_{\rm VLL, VRL}^{ij}$ is the sum of the box contribution and the Z penguin  one:
\begin{align}
	C_{\rm VLL}^{ij}
	&=-B_{\rm VLL}^{21ij}-\frac{\alpha_2}{4 \pi}Q_{ZL}^{(\nu)}P_{\rm ZL}^{21} \delta^{ij} \ ,\nonumber \\
	C_{\rm VRL}^{ij}
	&=-B_{\rm VRL}^{21ij}-\frac{\alpha_2}{4 \pi}Q_{ZL}^{(\nu)}P_{\rm ZR}^{21} \delta^{ij} \ ,
\end{align}
where the weak neutral-current coupling $Q_{ZL}^{(\nu)}=1/2$,
and $B_{\rm VL(R)L}^{21ij}$ and  $P_{\rm ZL(R)}^{21}$ denote
 the box  contribution  and the Z penguin contribution, respectively. 
 The $V$, $L$ and  $R$ denote the vector coupling, the left-handed one and the right-handed one,
 respectively.
In  addition to the W boson contribution,  there are the gluino $\tilde{g}$, 
the chargino $\chi^{\pm}$ and the neutralino $\chi^{0}$ mediated ones
%%%%%%%%%%%%%%%%%%%%%%%%%%%%%%%%%%
\footnote{The wino-higgsino mixing is tiny in our mass spectrum.}.
%%%%%%%%%%%%%%%%%%%%%%%%%%%%%%%%%%
We write each  contribution as follows:
\begin{align}
	B_{\rm VLL}^{sdij}
	&=B_{\rm VLL}^{sdij}(W)+B_{\rm VLL}^{sdij}(\chi^{\pm}) +B_{\rm VLL}^{sdij}(\chi^{0}) \ ,
  \nonumber \\
	B_{\rm VRL}^{sdij}
	&=B_{\rm VRL}^{sdij}(\chi^{\pm}) +B_{\rm VRL}^{sdij}(\chi^{0}) \ ,\nonumber \\
	P_{\rm ZL}^{sd}
	&=P_{\rm ZL}^{sd}(W) +P_{\rm ZL}^{sd}(H^{\pm}) +P_{\rm ZL}^{sd}(\tilde{g}) 
	+P_{\rm ZL}^{sd}(\chi^{\pm})+P_{\rm ZL}^{sd}(\chi^0)\ , \nonumber\\
	P_{\rm ZR}^{sd}
	&=P_{\rm ZR}^{sd}(\tilde{g}) +P_{\rm ZR}^{sd}(\chi^{\pm})+P_{\rm ZR}^{sd}(\chi^0)  \ ,
\label{boxpenguin}
\end{align}
where $(i,j)$ denotes the neutrinos  of final state.
 Explicit expressions are  given in Ref.\cite{GotoNote}.
It is well known that the most dominant contribution  comes from the Z penguin mediated chargino
 for the $K^+ \to \pi^+ \nu {\bar \nu}$ and $K_L \to \pi^0 \nu  {\bar \nu}$ decays 
\cite{Buras:2004uu}.

The branching ratio of $K^+ \to \pi^+ \nu {\bar \nu}$ and $K_L \to \pi^0 \nu  {\bar \nu}$  are obtained by replacing internal effect $F$ in Eqs. (\ref{BRKpPipnunu}) and 
(\ref{BRKLPi0nunu}) to $C_{\rm VLL}^{ij}+C_{\rm  VRL}^{ij}$  as follows:
\begin{align}
	BR(K^+ \to \pi^+ \nu {\bar \nu})
	&=\kappa \cdot  r_{K^+}
	\sum_{i=e, \mu, \tau} |C_{\rm VLL}^{ij}+C_{\rm VRL}^{ij}|^2 \ ,\\
	BR(K_L \to \pi^0 \nu  {\bar \nu})
	&=\kappa \cdot \frac{r_{K_L}}{r_{K^+}}\frac{\tau (K_L)}{\tau (K^+)} 
	\sum_{i=e, \mu, \tau} |{\rm Im} (C_{\rm VLL}^{ij}+C_{\rm VRL}^{ij})|^2 \  .
\end{align}

%%%%%%%%%%%%%%%%%%%%%%%%%%%%%%%%%%%%%%%%%%%%%%%%%%%%%%%%
\subsection{$\epsilon_K$ in the MSSM}
%%%%%%%%%%%%%%%%%%%%%%%%%%%%%%%%%%%%%%%%%%%%%%%%%%%%%%%%
It is well known that the CP violating parameter $\epsilon_K$ induced by the $K^0-{\bar K^0}$ oscillation gives us  one of the most serious constraint to the NP.
The general expression for  $\epsilon_K$ is given as
\begin{equation}
	\epsilon_K =
e^{i\phi_\epsilon}
%\kappa_{\epsilon} 
\sin{\phi_{\epsilon}} \left( \frac{\text{Im}(M_{12}^K)}{\Delta M_K}+ \xi\right) ,
\qquad \xi=\frac{\rm Im A_0}{\rm Re A_0},
\end{equation}
where $A_0$ is the $0$-isospin amplitude in the $K\rightarrow \pi\pi$ decay,
and 
%\begin{equation}
%\epsilon_K
%=\kappa_{\epsilon} \sin{\phi_{\epsilon}} \left( \frac{\text{Im}(M_{12}^K)}{\Delta M_K}
%+ \xi \right),  \qquad
%\xi
%=\frac{\text{Im} A_0}{\text{Re} A_0} , \qquad
%\phi_\epsilon
%=\tan^{-1}\left( \frac{2 \Delta M_K}{\Delta \Gamma_K} \right),
%\end{equation}
%with $A_0$ being the isospin zero amplitude in $K\to\pi\pi$ decays.
%Here, $M_{12}^K$ is the dispersive part of the  $K^0-\bar{K^0}$ mixing, $\Delta M_K$ is the mass difference in the neutral K meson.
$M_{12}^K$ is the dispersive part of the $K^0-{\bar K^0}$ oscillations, and  $\Delta M_K$ is the mass difference of the neutral $K$ meson.
The  effects of $\xi \ne 0$ and $\phi_{\epsilon} < \pi/4$
 were estimated  by Buras and Guadagnoli \cite{Buras:2008nn}.
% as: $\kappa_{\epsilon}=0.92 \pm 0.02$.
In the SM, the off-diagonal mixing amplitude $M_{12}^K$ is obtained as
\begin{align}
	M_K^{12}
	&=\langle K^0| \mathcal{H}_{\Delta S=2} |\bar{K^0} \rangle \nonumber \\
	&=\frac{4}{3}\left( \frac{G_F}{4 \pi} \right)^2 M_W^2 \hat{B}_K F_K^2 M_K 
	\Big[ \eta_{cc} (V_{cs}V_{cd}^*)^2 S(x_c)+\eta_{tt} (V_{ts}V_{td}^*)^2 S(x_t) \\
	&~~~~~~~~~~~~~~~~~~~~~+2 \eta_{ct} (V_{cs}V_{cd}^*) (V_{ts}V_{td}^*) S(x_c,x_t) \Big] , 
\nonumber
\label{m12K}
\end{align}
where $S(x)$ denotes the SM one-loop functions \cite{Inami:1980fz}, 
and $\eta_{cc,tt,ct}$ are the QCD corrections \cite{Buras:2008nn}.
 Recent  lattice calculations give us the precise determination of the 
 $\hat{B}_K$ parameter \cite{Bae:2013lja,Aoki:2013ldr}.

Once taking account of the NP effect,  the expression of $M_{12}^K$ is modified.
In the case of the SUSY, new contributions to the box diagrams are given 
 by  the gluino $\tilde{g}$, the charged Higgs $H^{\pm}$, the chargino $\chi^{\pm}$ and the neutralino $\chi^{0}$ exchanges:
\begin{align}
	M^K_{12}
	&=M_{12}^{K, {\rm SM}} + M_{12}^{K, {\rm SUSY}} \nonumber \\
	&=M_{12}^{K} (W) + M_{12}^{K} (H^{\pm})+ M_{12}^{K} (\chi^{\pm}) + M_{12}^{K} (\chi^{0})+ M_{12}^{K} (\tilde{g})+ M_{12}^{K} (\chi^{0}\tilde{g}) .  \nonumber
\end{align}
The explicit formula has been presented in Ref. \cite{GotoNote}.

%%%%%%%%%%%%%%%%%%%%%%%%%%%%%%%%%%%%%%%%%%%%%%%%%%%%%%%%%%%
%%%%%%%%%%%%%%%%%%%%%%%%%%%%%%%%%%%%%%%%%%%%%%%%%%%%%%%%%%%
\section{Setup of the squark flavor mixing}
%%%%%%%%%%%%%%%%%%%%%%%%%%%%%%%%%%%%%%%%%%%%%%%%%%%%%%%%%%%
%%%%%%%%%%%%%%%%%%%%%%%%%%%%%%%%%%%%%%%%%%%%%%%%%%%%%%%%%%%
\label{sec:Setup}
We present the setup of our calculation in the framework of the high-scale SUSY.
Recent LHC results for the SUSY search may suggest the high-scale SUSY, ${\mathcal O}(10-1000)$ TeV
\cite{Altmannshofer:2013lfa,Moroi:2013sfa,Tanimoto:2014eva,Hisano:2013cqa}
since the lower bounds of the gluino mass and squark masses exceed 1 TeV.
Taking account of these recent results, we consider the possibility of the high-scale SUSY 
at $10,~50 ~{\rm TeV}$, in which 
the $K \to \pi \nu {\bar \nu}$ decays  and $\epsilon_K$ are discussed..

Another important experimental result should be mentioned is the Higgs discovery.
The Higgs mass $m_H \simeq 126$ GeV gives effect to the  SUSY mass spectrum.
In general, there are two possibility to get Higgs mass value, one is the heavy stop around 10 TeV, and the another is the large  $X_t = A_0 - \mu\cot \beta$ given by the  A-term.
In the case that the SUSY scale is $10$ to $50$ TeV, we have already obtained the SUSY mass spectra which realize the Higgs mass at the electroweak scale with Renormalization Group Equation (RGE) running in  previous work \cite{Tanimoto:2014eva}.
We use this numerical  result for the SUSY particle mass spectrum. 
In this study,  the 1st and 2nd squark are almost degenerated
due to the assumption of the universal soft masses.
On the other hand, the 3rd squark mass obtains the large contribution from the RGE running  due to the large Yukawa coupling of the top-quark.
Therefore,  the mixing between 1st and 2nd is negligible, and it is taken account in the subsequent discussion for squark flavor mixing. 
The SUSY spectra at $10$ and  $50$ TeV are given in  Appendix A.

%In order to compare with the case of TeV scale SUSY, we also consider  the  case of  $2$TeV
% for the SUSY mass.
%Then, the Higgs mass is realized with  ${\rm tan} \beta=20$, $m_{\tilde t} =2 ~{\rm TeV}$ and 
%$X_t \simeq 2~{\rm TeV}$.  Then, 
%we also set the gaugino masses $M_1\simeq 0.5$TeV, $M_2\simeq 1$TeV and  
%$M_3\simeq 3$TeV,  which satisfy the direct search bound and the well-known mass ratio,
% $M_3 : M_2 : M_1 \approx 6:2:1$.

Once the SUSY mass spectrum is fixed, 
we can calculate the left-right mixing angle $\theta^q$, which is defined as
\begin{equation}
	\theta^{d}
	\simeq \frac{m_b(A_0 - \mu \tan \beta)}{m_{\tilde b_L}^2-m_{\tilde b_R}^2} ,
	\qquad\qquad
	\theta^{u}
	\simeq \frac{m_t(A_0 - \mu \cot \beta)}{m_{\tilde t_L}^2-m_{\tilde t_R}^2} .
\end{equation}
In the case of the SUSY scale to be $10$ and $50$TeV, the left-right mixing angles
of squarks and sleptons  are
very small as 
($\theta^{d}\sim 0.0062$, $\theta^{u}\sim 0.0024$, $\theta^{e}\sim 0.014$) and
 ($\theta^{d}\sim 0.0009$, $\theta^{u}\sim 0.0007$, $\theta^{e}\sim 0.005$), respectively.

The SUSY brings the new  flavor mixing through
the quark-squark-gaugino couplings and the lepton-slepton-gaugino ones.
The $6\times 6$ squark mass matrix  $M_{\tilde q}^2$ in the super-CKM basis turns to the mass eigenstate basis by diagonalizing with rotation matrix $\Gamma ^{(q)}$ as  
\begin{equation}
m_{\tilde q}^2=\Gamma^{(q)}M_{\tilde q}^2 \Gamma^{(q)  \dagger} \ ,
\end{equation}
where $\Gamma^{(q)}$ is the $6\times 6$ unitary matrix, and we decompose it 
into the $3\times 6$ matrices as $\Gamma ^{(q)}=(\Gamma _{L}^{(q)}, \ \Gamma _{R}^{(q))})^T$ in the following expressions:
\begin{align}
	\Gamma _{L}^{(q)}&=
	\begin{pmatrix}
	c_{13}^{qL} & 0 & s_{13}^{qL} e^{-i\phi_{13}^{qL}} c_{\theta^q} & 0 & 0 & -s_{13}^{qL} e^{-i\phi_{13}^{qL}} s_{\theta^q} e^{i \phi^q} \\
	-s_{23}^{qL} s_{13}^{qL} e^{i(\phi_{13}^{qL}-\phi_{23}^{qL})} & c_{23}^{qL} & s_{23}^{qL}c_{13}^{qL}e^{-i\phi_{23}^{qL}}	c_{\theta^q} & 0 & 0 & -s_{23}^{qL}c_{13}^{qL}e^{-i\phi_{23}^{qL}}s_{\theta^q}e^{i\phi^q} \\
	-s_{13}^{qL}c_{23}^{qL}e^{i\phi_{13}^{qL}} & -s_{23}^{qL}e^{i\phi_{23}^{qL}} &c_{13}^{qL}c_{23}^{qL}c_{\theta^q} & 0 & 0 &  -c_{13}^{qL}c_{23}^{qL}s_{\theta^q} e^{i\phi^q}
\end{pmatrix}, \nonumber \\
\nonumber\\
\Gamma _{R}^{(q)}&=
\begin{pmatrix}
	0 & 0 & s_{13}^{qR} s_{\theta^q} e^{-i \phi_{13}^{qR}} e^{-i\phi^q } & c_{13}^{qR} & 0 & s_{13}^{qR} e^{-i \phi_{13}^{qR}} c_{\theta^q} \\
	0 & 0 & s_{23}^{qR} c_{13}^{qR} s_{\theta^q} e^{-i \phi_{23}^{qR}} e^{-i\phi^q } & -s_{13}^{qR} s_{23}^{qR} e^{i(\phi_{13}^{qR}-\phi_{23}^{qR})} & c_{23}^{qR} & s_{23}^{qR} c_{13}^{qR} e^{-i \phi_{23}^{qR}} c_{\theta^q} \\
0 & 0 & c_{13}^{qR} c_{23}^{qR} s_{\theta^q} e^{-i\phi^q} & -s_{13}^{qR} c_{23}^{qR} e^{i\phi_{13}^{qR}} & -s_{23}^{qR} e^{i \phi_{23}^{qR}} & 
c_{13}^{qR} c_{23}^{qR} c_{\theta^q} \\
\end{pmatrix},
\label{mixing}
\end{align} 
where we use abbreviations $c_{ij}^{qL,qR}=\cos\theta_{ij}^{qL,qR}$, $s_{ij}^{qL,qR}=\sin\theta_{ij}^{qL,qR}$,  $c_{\theta^q} =\cos \theta^q$ and $s_{\theta^q} =\sin \theta^q$.
It is remarked that we take $s_{12}^{qL,qR}=0$ due to the degenerate squark masses of the 1st and the 2nd  families as noted in Appendix A.
The angle $\theta^q$ is  the left-right mixing angle 
 between  $\tilde q_{L}$ and $\tilde q_{R}$,
 and they are calculable as mentioned above.
Then, there are free mixing parameters $\theta_{ij}^{qL,qR}$ and $\phi_{ij}^{qL,qR}$.
For simplicity, we assume $s_{ij}^{qL}=s_{ij}^{qR}$. 
On the other hand, we scatter $\phi_{ij}^{qL}$ and $\phi_{ij}^{qR}$ 
in the $0\sim 2\pi$ range independently.
 It should be  noted that the mixing angles $s_{ij}^{qL(R)}$ have not been constrained by
the experimental data of B, D and K mesons in the framework of the high-scale SUSY 
\cite{Tanimoto:2014eva}.

For the lepton sector,
the mixing matrices $\Gamma _{L(R)}^{(\ell)}$ have the  same structure as the quark one in 
 the charged-lepton flavors,
 however,  there is only  the left-handed $\Gamma _{L}^{(\nu)}$  in neutrinos.
 
 As well known,  the charged Higgs and the chargino contributions dominate
the  $K \to \pi \nu{\bar \nu}$ processes  \cite{Buras:2004uu}.
Since the SUSY scale is high in our scheme, the charged Higgs are heavy, ${\cal O}(10{\rm TeV})$,
so the charged Higgs contribution is  suppressed in our framework.
On the other hand, the dominant SUSY contribution to $\epsilon_K$  comes from 
 the gluino box diagram if the flavor mixing angles of 
 the down-squark and the up-squark  are comparable. In addition,  the chargino box diagram 
is also non-negligible. 
 Consequently, we will discuss the both cases in which the down-squark mixing angles  $s_{ij}^{dL(R)}$ are negligible small and  are comparable to  the up-squark mixing angles
 $s_{ij}^{uL(R)}$.
We scan the phases of  Eq.(\ref{mixing}) for up-squarks, down-squarks, charged-sleptons and sneutrinos in the region of $0 \sim 2\pi$ independently.  

In our framework,  the $K \to \pi \nu{\bar \nu}$ processes
are dominated by the Z penguin mediated the chargino exchange, 
$P_{\rm ZL}^{sd}(\chi^{\pm})$  in Eq.(\ref{boxpenguin}) , which
are occurred through the $\tilde t_L s_L(d_L)\chi^\pm$ and $\tilde t_R s_L(d_L)\chi^\pm$
interactions, respectively. 
In our basis, the relevant mixing is given by
\begin{equation}
(\Gamma _{CL}^{(d)})_I^{\alpha q}\equiv 
(\Gamma _{L}^{(u)}  V_{\rm CKM})_I^q (U_+)^\alpha_1
+\frac{1}{g_2}(\Gamma _{R}^{(u)} \hat f_U V_{\rm CKM})_I^q (U_+)^\alpha_2 \ ,
\end{equation}  
where $q=s, d$,  $I=1-6$ for up-squarks,  and $\alpha=1,2$  for charginos.
 The  $V_{\rm CKM}$ is the CKM matrix, 
and   $U_+$ is the  $2\times 2$ unitary matrix which diagonalize 
$M^\dagger_C M_C$, where $M_C$ is the $2\times 2$ chargino mass matrix.
The $\hat f_U$ denotes the yukawa coupling defined by $\hat f_U v \sin{\beta}=
{\rm diag}(m_u,m_c,m_t)$.
Therefore,  the combinations of  mixing angles and phases 
in Eq.(\ref{mixing}), $c_{13}^{qL}s_{13}^{qL}s_{23}^{qL}  e^{i(\phi_{13}^{qL}-\phi_{23}^{qL})}$
and $c_{13}^{qR}s_{13}^{qR}s_{23}^{qR}  e^{i(\phi_{13}^{qR}-\phi_{23}^{qR})}$ are 
important for our numerical analyses in the next section.
We show the formula  for $P_{\rm ZL}^{sd}(\chi^{\pm})$  in Appendix B.

%%%%%%%%%%%%%%%%%%%%%%%%%%%%%%%%%%%%%%%%%%%%%%%%%%%%%%%%%%%
%%%%%%%%%%%%%%%%%%%%%%%%%%%%%%%%%%%%%%%%%%%%%%%%%%%%%%%%%%%
\section{Numerical analysis}
%%%%%%%%%%%%%%%%%%%%%%%%%%%%%%%%%%%%%%%%%%%%%%%%%%%%%%%%%%%
%%%%%%%%%%%%%%%%%%%%%%%%%%%%%%%%%%%%%%%%%%%%%%%%%%%%%%%%%%%
\label{sec:Numerical}
Let us discuss the high-scale SUSY contribution to the  $K \to \pi \nu{\bar \nu}$ processes
 by correlating  with $\epsilon_K$ \cite{Blanke:2009pq}. 
At present, we cannot confirm whether the SM prediction $\epsilon_K^{\rm SM}$
is  in agreement with the experimental value $\epsilon_K^{\rm exp}$ 
because  there remains the theoretical  uncertainty with an order of a few ten percent. 
However, the theoretical uncertainties of $\epsilon_K$ are expected to be reduced significantly
 in the near future.  Actually, the lattice calculations of $\hat B_K$  will be improved
 significantly \cite{Bae:2013lja,Aoki:2013ldr}, whereas   $|V_{cb}|$ and 
the CKM phase $\gamma$ will be measured more precisely in Belle-II.
Therefore, we will be able to test the correlation between $K \to \pi \nu{\bar \nu}$ and  $\epsilon_K$.
%We consider two cases of $\epsilon_K$,
%\begin{itemize}
%	\item {\bf Case (I)} ~~~Small SUSY effects in $\epsilon_K$ 
%	: $\left| \epsilon_K^{\rm SUSY}/\epsilon_K^{\rm SM+SUSY} \right| \leq  5 \%$,
%	\item {\bf Case (II)} ~~Large(Sizable) SUSY effects in $\epsilon_K$ 
%	: $\left| \epsilon_K^{\rm SUSY}/\epsilon_K^{\rm SM+SUSY} \right| \geq 20 \%$.
%\end{itemize}
%In the Case (I), we set the SUSY contribution to $\epsilon_K$ is at most $\pm 5\%$.
%On the other hand, in the Case (II) we assume the  SUSY contribution to $\epsilon_K$
% larger than $20 \%$.
 
In our previous work, we have examined the sensitivity of the high-scale SUSY with $10$ 
and $50$ TeV to $\epsilon_K$.
  It is found that the SUSY contribution to $\epsilon_K$  is allowed up to 
 $40 \%$. We begin to discuss the SUSY contribution at the $10$ TeV scale.
The present uncertainties in the SM prediction for $\epsilon_K$  are due to  
the CKM elements  $V_{cb}$, $\bar\rho$ and $\bar\eta$, and the $\hat B_K$ parameter.
We take the CKM parameters $V_{cb}$,   $\bar\rho$ and $\bar\eta$ at the 90 \% C.L.
of the  experimental data:
%%%%%%%%%%%%%%%%%%%%%%%%%%%%%%%%%%%%%%%%
\begin{equation}
|V_{cb}|=(41.1 \pm 1.3)\times 10^{-3}, \qquad  \bar\rho=0.117\pm 0.021,  \qquad  
\bar\eta= 0.353\pm 0.013 .
\end{equation}
%%%%%%%%%%%%%%%%%%%%%%%%%%%%%%%%%%%%%%%%
For the  $\hat B_K$ parameter,  the recent result of the  lattice calculations 
is given as
\cite{Bae:2013lja,Aoki:2013ldr};
\begin{align}
\hat{B}_K=0.766\pm 0.010 \ ,	
\label{BK}
\end{align}
 which is used with the error-bar of  90\% C.L. in our calculation.
%%%%%%%%%%%%%%%%%%%%%%%%%%%%%%%%%%%%%%%%%%%%%%%%%%%%%%%%%%%
%\begin{wrapfigure}{r}{8cm}
%         		\begin{center}
%		         \includegraphics[width=8cm]{KL-Kp_10TeV_sdsu01.eps}
%		  \caption{The predicted $BR(K_L \to \pi^0 \nu  {\bar \nu})$ versus $BR(K^+ \to \pi^+ \nu %{\bar \nu})$at  the SUSY scale of $10$ TeV  with the mixing angle of $s^u=s^d=0.1$. 
% The pink cross denotes the SM predictions.
%The red dashed lines are the $1\sigma$ experimental values for $BR(K^+ \to \pi^+ \nu {\bar \nu})$. %The green slanting line shows the Grossman-Nir bound \cite{Grossman:1997sk}.} 
%		         \label{fig:1}
%	         \end{center}
%        \end{wrapfigure} 
%%%%%%%%%%%%%%%%%%%%%%%%%%%%%%%%%%%%%%%%%%%%%%%%%%%%%%%%%%%
%%%%%%%%%%%%%%%%%%%%%%%%%%%%%%%%%%%%%%%%%%%%%%%%%%%%%%%%%%%%%
\begin{figure}[t]
\begin{center}
\includegraphics[width=10cm]{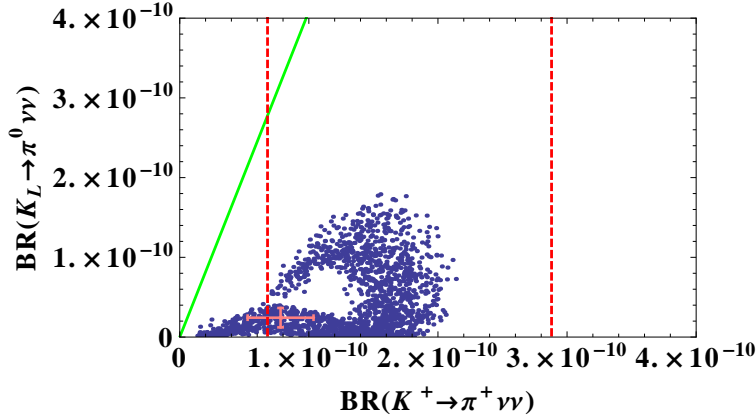}
\end{center}
\caption{The predicted $BR(K_L \to \pi^0 \nu  {\bar \nu})$ versus
 $BR(K^+ \to \pi^+ \nu {\bar \nu})$ at  the SUSY scale of $10$ TeV  with the mixing angle of $s^u=s^d=0.1$. 
 The pink cross denotes the SM predictions.
The red dashed lines are the $1\sigma$ experimental bounds for $BR(K^+ \to \pi^+ \nu {\bar \nu})$. The green slanting line shows the Grossman-Nir bound.} 
\label{fig:1}
\end{figure}
%%%%%%%%%%%%%%%%%%%%%%%%%%%%%%%%%%%%%%%%%%%%%%%%%%%%%%%%%%%%

In the beginning, we show the numerical  results at the SUSY scale of $10$ TeV. 
Fig.\ref{fig:1} shows the predictions on the  $BR(K_L \to \pi^0 \nu  {\bar \nu})$ vs. 
$BR(K^+ \to \pi^+ \nu {\bar \nu})$ plane, where phase parameters are constrained by
the observed $|\epsilon_K|$ with the experimental error-bar of $90\%$C.L. 
Here, we fix the mixing parameters
 in Eq.(\ref{mixing}) by taking the common value $s_{i3}^{uL}=s_{i3}^{uR}=s^u=0.1 \ (i=1,2)$
and  $s_{i3}^{dL}=s_{i3}^{dR}=s^d=0.1 \ (i=1,2)$ for the up-quark and the down-quark sectors, respectively.
The Z penguin mediated chargino  dominates the SUSY contribution to these branching ratios.

The SUSY contributions can enhance the  branching ratio
of $K_L \to \pi^0 \nu {\bar \nu}$
in eight times compared with the SM predictions in Eq.(\ref{SMprediction}), $1.8\times 10^{-10}$
although it is much smaller than the  Grossman-Nir bound.
On the other hand,  the predicted $BR(K^+ \to \pi^+ \nu {\bar \nu})$
increases up to three times, $2.1\times 10^{-10}$.
%It is understandable because $K^+ \to \pi^+ \nu {\bar \nu}$ is the CP-concerving process and not so sensitive to the CP violating phase. 
%In contrast, $K_L \to \pi^0 \nu  {\bar \nu}$ only depends on imaginary part as shown 
%in Eq.(\ref{BRKLPi0nunu}) and is correlated with the $\epsilon_K$ considerably.
It is also noticed that the predicted region of $BR(K_L\to\pi^0 \nu {\bar \nu})$ is reduced
 to much  smaller than  $10^{-11}$ due to the cancellation between the SM and SUSY contributions.
The $BR(K^+ \to \pi^+ \nu {\bar \nu})$ could be reduced to $1.3\times 10^{-11}$.

%%%%%%%%%%%%%%%%%%%%%%%%%%%%%%%%%%%%%%%%%%%%%%%%%%%%%%%%%%%%%
\begin{figure}[t]
\includegraphics[width=8cm]{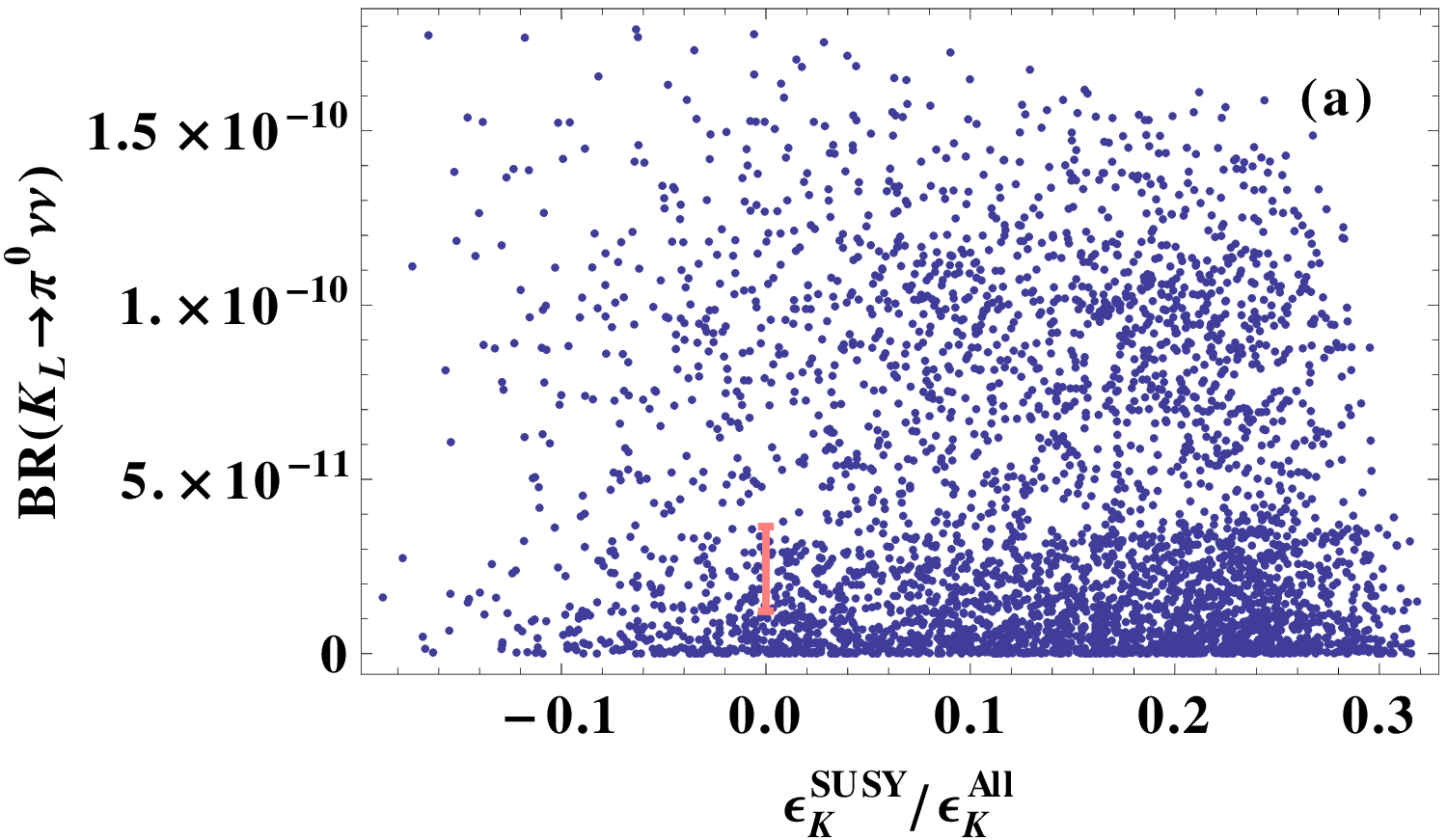}
\hspace{0.3cm}
\includegraphics[width=8cm]{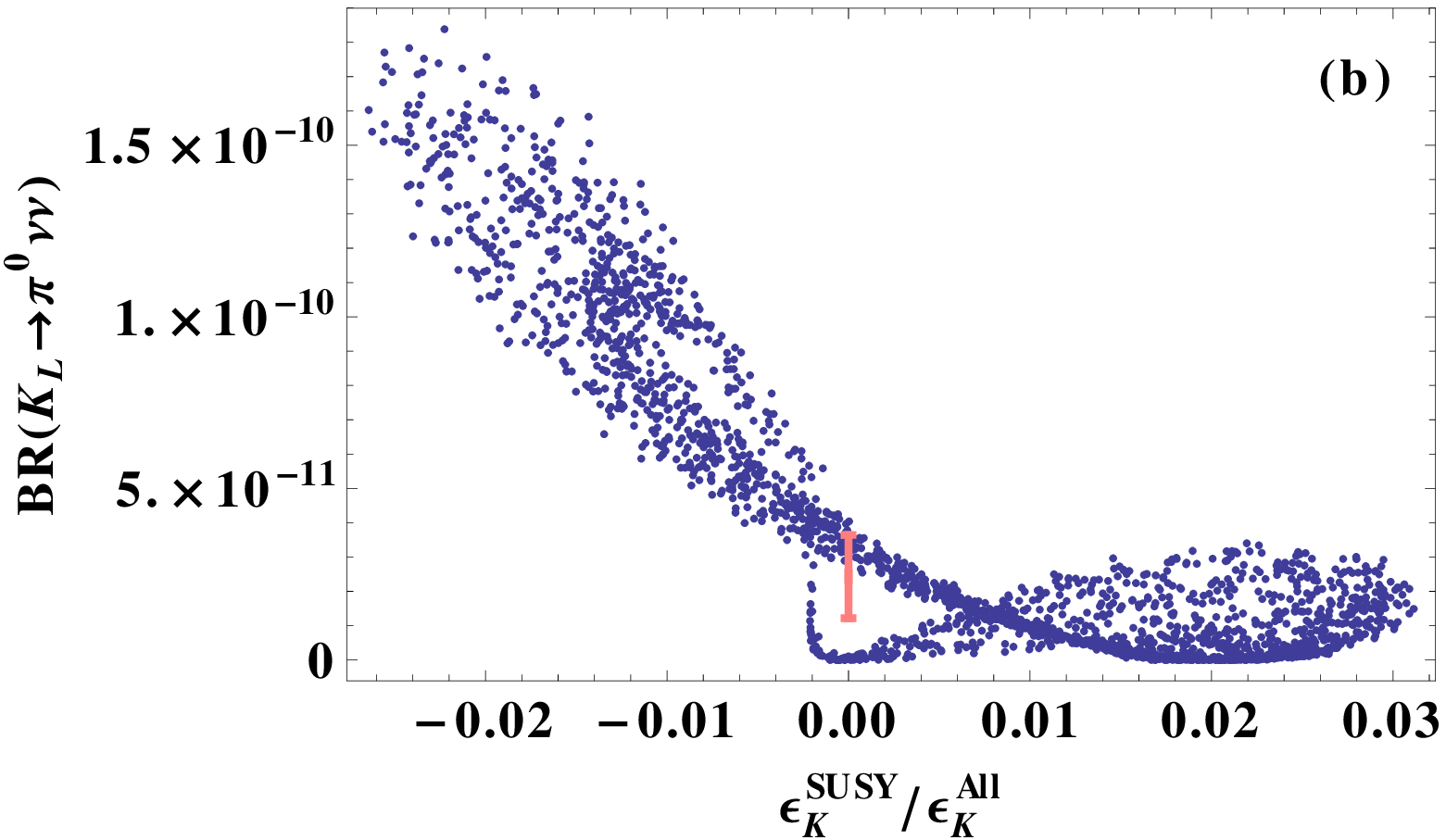}
\hspace{1cm}
\caption{The predicted $BR(K_L \to \pi^0 \nu  {\bar \nu})$ versus the SUSY contribution ratio 
of $\epsilon_K$  at the SUSY scale of $10$ TeV in the case of
(a) $s^u=s^d=0.1$ and (b) $s^u=0.1, \ s^d=0$.  The pink short line denotes
the SM prediction  with the error-bar of 90\%C.L.} 
\label{fig:2}
\end{figure}
%%%%%%%%%%%%%%%%%%%%%%%%%%%%%%%%%%%%%%%%%%%%%%%%%%%%%%%%%%%%
%%%%%%%%%%%%%%%%%%%%%%%%%%%%%%%%%%%%%%%%%%%%%%%%%
%%%%%%%%%%%%%%%%%%%%%%%%%%%%%%%%%%%%%%%%%%%%%%%%%%%%%%%%%%%
\begin{figure}[t]
\includegraphics[width=8cm]{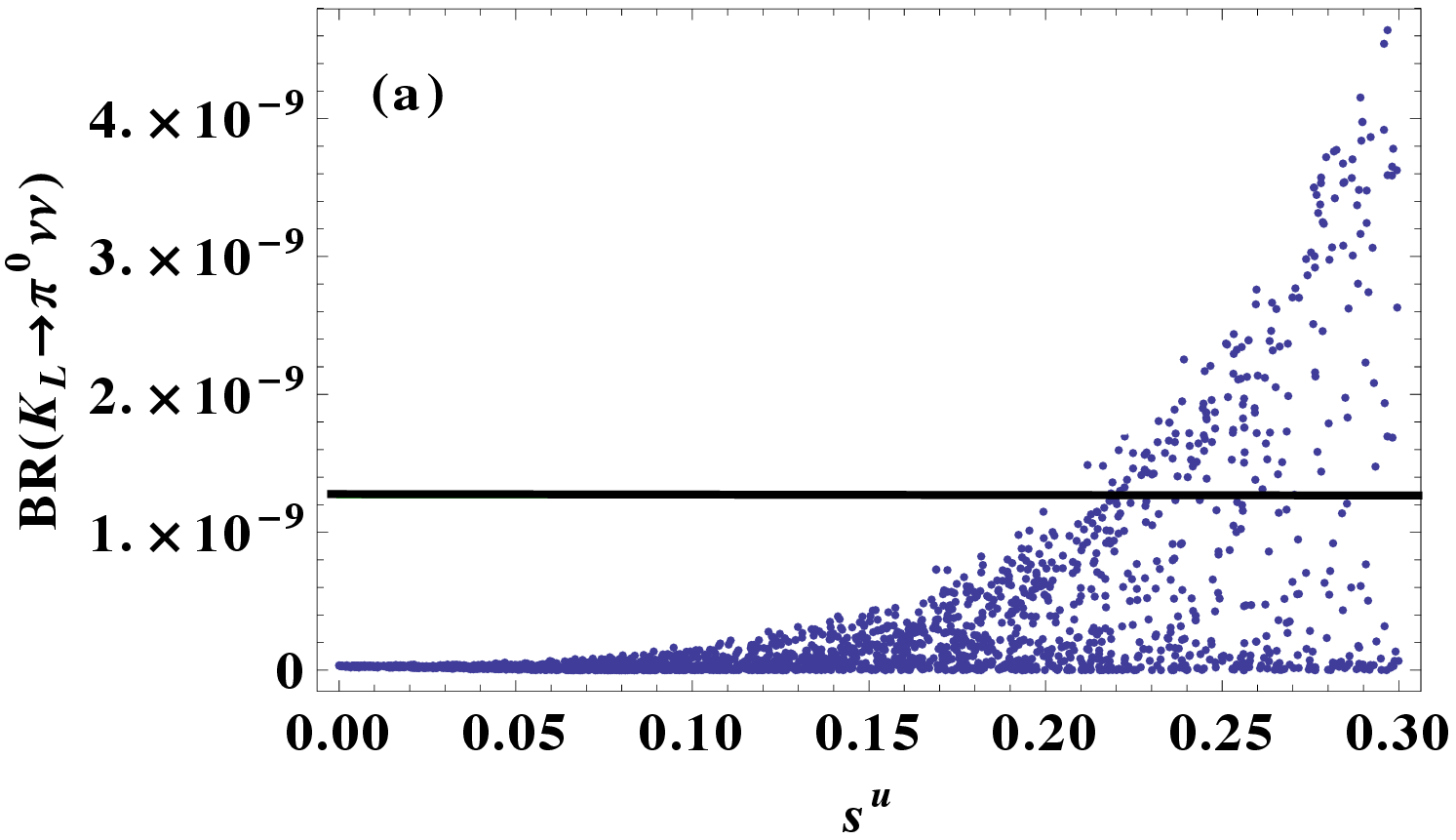}
\hspace{0.5cm}
\includegraphics[width=8cm]{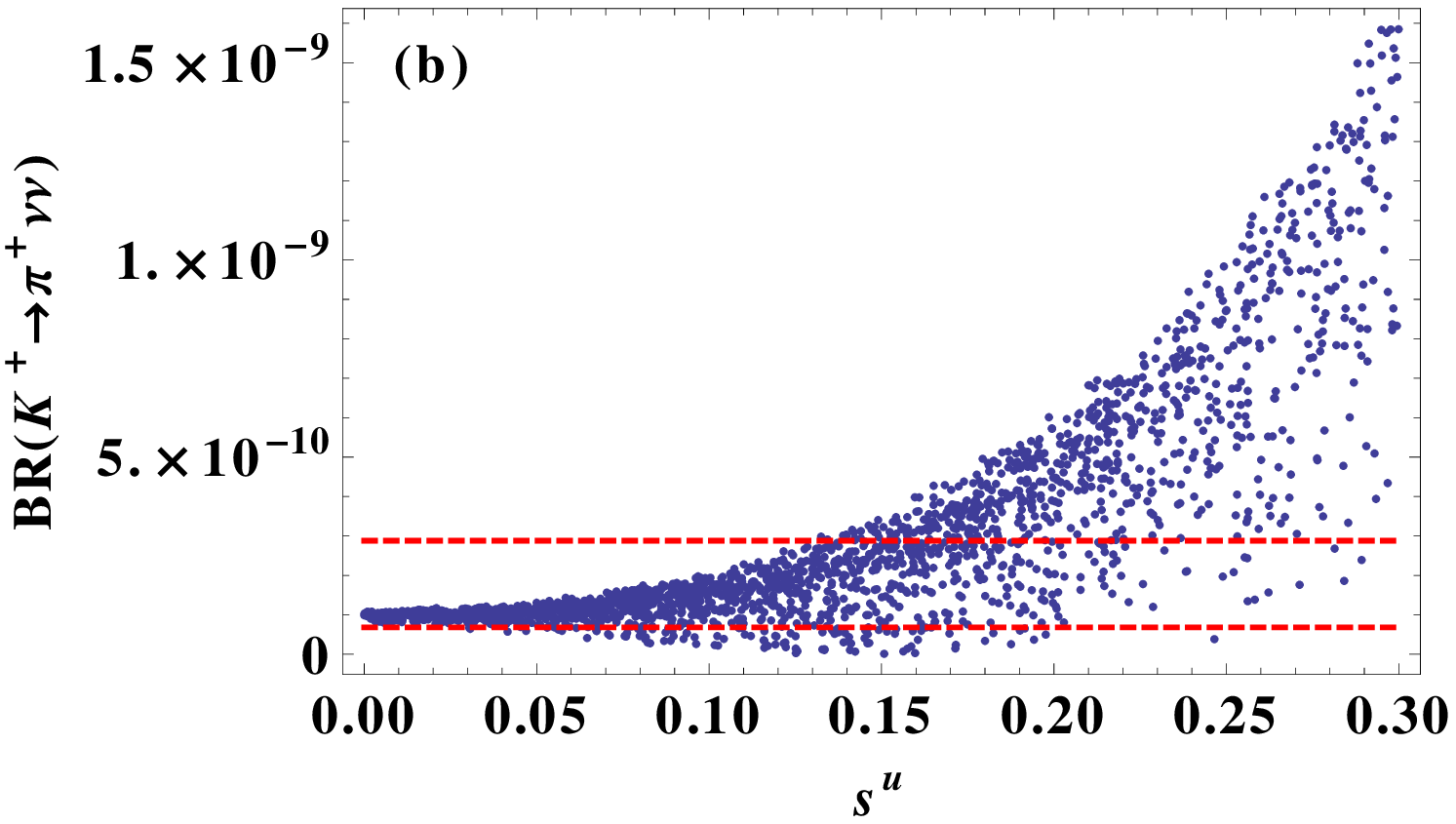}
\caption{The predicted $s^u$ dependence of (a) $BR(K_L \to \pi^0 \nu  {\bar \nu})$ and 
(b) $BR(K^+ \to \pi^+ \nu {\bar \nu})$ at  the SUSY scale of $10$ TeV, where $s^d$ is scanned in the region of $0\sim 0.3$ independent of $s^u$.
The red dashed lines denote the $1\sigma$ experimental bounds for $BR(K^+ \to \pi^+ \nu {\bar \nu})$.  The black line corresponds to  the Grossman-Nir bound together with the experimental upper bound of $BR(K^+ \to \pi^+ \nu {\bar \nu})$ with $3\sigma$.} 
\label{fig:3}
\end{figure}
%%%%%%%%%%%%%%%%%%%%%%%%%%%%%%%%%%%%%%%%%%%%%%%%%%%%%%%%%%%%

%This situation is understandable as follows.
%%%%%%%%%%%%%%%%%%%%%%%%%%%%%%%%%%%%%%%%%%%%%%%%%
% Since the phase tuning is needed to realize the large SUSY effect in $\epsilon_K$, the allowed %region of the $BR(K_L \to \pi^0 \nu  {\bar \nu})$ become small as the SUSY contribution in the %$\epsilon_K$ gets large.
%Hence case (II) gives severe constraints compared with case (I).  
%Therefore, the difference of SUSY effect in $\epsilon_K$ has effect to $K_L\to\pi^0\nu {\bar \nu}$.
%%%%%%%%%%%%%%%%%%%%%%%%%%%%%%%%%%%%%%%%%%%%%%%%%
%In conclusion, it is emphasized that even if the SUSY contribution is tiny in 
%$\epsilon_K$, and so it is in agreement  with the SM prediction in the near future,
%the branching  ratios  of $K_L \to \pi^0 \nu  {\bar \nu}$ 
%and $K^+ \to \pi^+ \nu {\bar \nu}$ could be deviated from the SM significantly.
%%%%%%%%%%%%%%%%%%%%%%%%%%%%%%%%%%%%%%%%%%%%%%%%%

We discuss the  correlation between  $\epsilon_K$ and  $BR(K_L \to \pi^0 \nu  {\bar \nu})$ 
in Fig. \ref{fig:2},
in which  (a)  $s^u=s^d=0.1$ and (b) $s^u=0.1, \ s^d=0$.
The transverse axis denotes the SUSY contribution in  $|\epsilon_K|$.
If the down-squark mixing $s^d$ is comparable to the up-squark mixing   $s^u$,
there is no correlation between them as seen in  Fig. \ref{fig:2}(a),
where the Z penguin mediated chargino  dominates
the  SUSY contribution of   $K_L \to \pi^0 \nu  {\bar \nu}$, and  
 the gluino box diagram
dominates the SUSY contribution of   $\epsilon_K$.
The gluino contribution of $30\%$ is possible in  $\epsilon_K$.

On the other hand, 
if the down-squark mixing $s^d$ is tiny compared with  the up-squark mixing   $s^u$,
the Z penguin mediated chargino  dominates both
  SUSY contributions of   $K_L \to \pi^0 \nu  {\bar \nu}$ and    $\epsilon_K$.
Then, it is found  a correlation between them as seen in  Fig. \ref{fig:2}(b),
where the chargino contribution  to  $\epsilon_K$ is at most $3\%$.
This correlation is due to the  difference  of the phase structure
  between the penguin diagram and the box diagram of the chargino.

In conclusion,  $\epsilon_K$ could be  deviated  from the SM prediction
 in ${\cal O}(10\%)$ due to the gluino box diagram, 
whereas the Z penguin mediated chargino  could enhance the branching ratio of
$K_L \to \pi^0 \nu  {\bar \nu}$ from the SM prediction.
%It is easily seen that the phase factors are  the different in 
% in the chargino  box diagram and the Z penguin mediated the chargino.
%When one considers  the stop exchange diagrams, the chargino box  leads to  the phase factor 
%$\exp [2i(\phi^{uL(R)}_{13}-\phi^{uL(R)}_{23})]$,
%while the Z penguin mediated the chargino gives  $\exp [i(\phi^{uL(R)}_{13}-\phi^{uL(R)}_{23})]$.
%For  $\phi^{uL(R)}_{13}-\phi^{uL(R)}_{23}=\pm\pi/2$,
%the imaginary part of the   $K_L \to \pi^0 \nu  {\bar \nu}$ amplitude is maximal,
% but the imaginary part of the chargino box vanishes. Thus, the SUSY contribution
% to $\epsilon_K$ is suppressed.

%%%%%%%%%%%%%%%%%%%%%%%%%%%%%%%%%%%%%%%%%%%%%%%%%
Next, in order to see the  mixing angle $s^u$ dependence of the branching ratios,
 we plot the predicted regions on $BR(K_L \to \pi^0 \nu  {\bar \nu})$ vs. $s^u$ 
and $BR(K^+ \to \pi^+ \nu {\bar \nu})$ vs. $s^u$ planes
  taking $s^u=0 \sim 0.3$
in Fig.\ref{fig:3} (a) and (b).
 We scan  $s^d$ in the region of $0\sim 0.3$ independent of $s^u$
 although the gluino contribution is much suppressed compared with the chargino one.
In this plot, the SUSY contribution to  $\epsilon_K$ is  free ($0-40\%$), but 
the experimental  constraint of $|\epsilon_K|$ with the error-bar of 90\%C.L. is taken account.
We show  the upper bound given by the Grossman-Nir bound together with  the experimental upper bound
of $BR(K^+ \to \pi^+ \nu {\bar \nu})$ with $3\sigma$ by the black line, at which
the predicted $BR(K_L \to \pi^0 \nu  {\bar \nu})$ should be cut.
  Namely, the observed upper bound of $BR(K^+ \to \pi^+ \nu {\bar \nu})$
 gives the  constraint for the predicted $BR(K_L \to \pi^0 \nu  {\bar \nu})$
at $s^u$ larger than $0.2$. 
The precise   experimental measurement of $BR(K^+ \to \pi^+ \nu {\bar \nu})$
will lower  the predicted upper bound of $BR(K_L \to \pi^0 \nu  {\bar \nu})$.

%%%%%%%%%%%%%%%%%%%%%%%%%%%%%%%%%%%%%%%%%%%%%
%%%%%%%%%%%%%%%%%%%%%%%%%%%%%%%%%%%%%%%%%%%%%

\begin{figure}[t]
\begin{center}
\includegraphics[width=10cm]{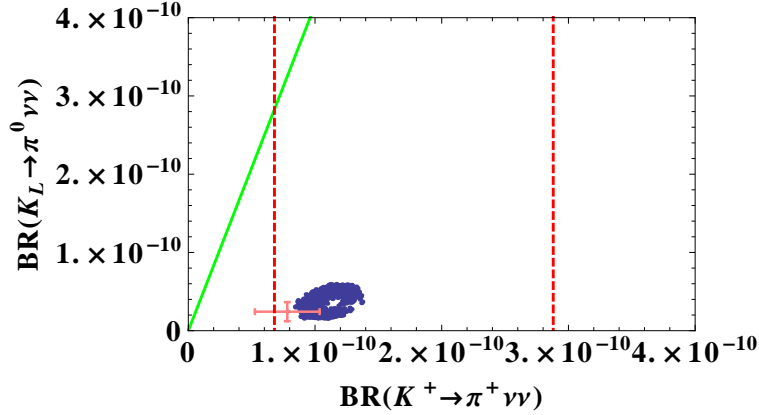}
\end{center}
\caption{The predicted $BR(K_L \to \pi^0 \nu  {\bar \nu})$ versus
 $BR(K^+ \to \pi^+ \nu {\bar \nu})$ at  the SUSY scale of $50$ TeV  with the mixing angle of $s^u=s^d=0.3$. 
 The pink cross denotes the SM predictions.
The red dashed lines are the $1\sigma$ experimental values for $BR(K^+ \to \pi^+ \nu {\bar \nu})$. The green slanting line shows the Grossman-Nir bound \cite{Grossman:1997sk}.} 
\label{fig:4}
\end{figure}

Let us  discuss the case of the SUSY scale of $50$ TeV.
Fig. \ref{fig:4} shows the predictions on the  $BR(K_L \to \pi^0 \nu  {\bar \nu})$ and 
$BR(K^+ \to \pi^+ \nu {\bar \nu})$ plane at the  SUSY scale of $50$ TeV, where the mixing angle 
is fixed at  $s^u=s^d=0.3$.
 Although the predicted region is  reduced considerably 
 comparing to the case of the 10 TeV  scale  in Fig. \ref{fig:1},
the predicted branching ratio of $K_L \to \pi^0 \nu  {\bar \nu}$ is  enhanced
 in two times from the SM prediction, and
the branching ratio of $K^+ \to \pi^+ \nu {\bar \nu}$ could be enhanced from the SM prediction
 in  three times.

%%%%%%%%%%%%%%%%%%%%%%%%%%%%%%%%%%%%%%%%%%%%%%%%%%%%%%%%%%%
%%%%%%%%%%%%%%%%%%%%%%%%%%%%%%%%%%%%%%%%%%%%%%%%%%%%%%%%%%%%
%%%%%%%%%%%%%%%%%%%%%%%%%%%%%%%%%%%%%%%%%%%%%%%%%%%%%%%%%%%
\begin{figure}[h]
\includegraphics[width=8cm]{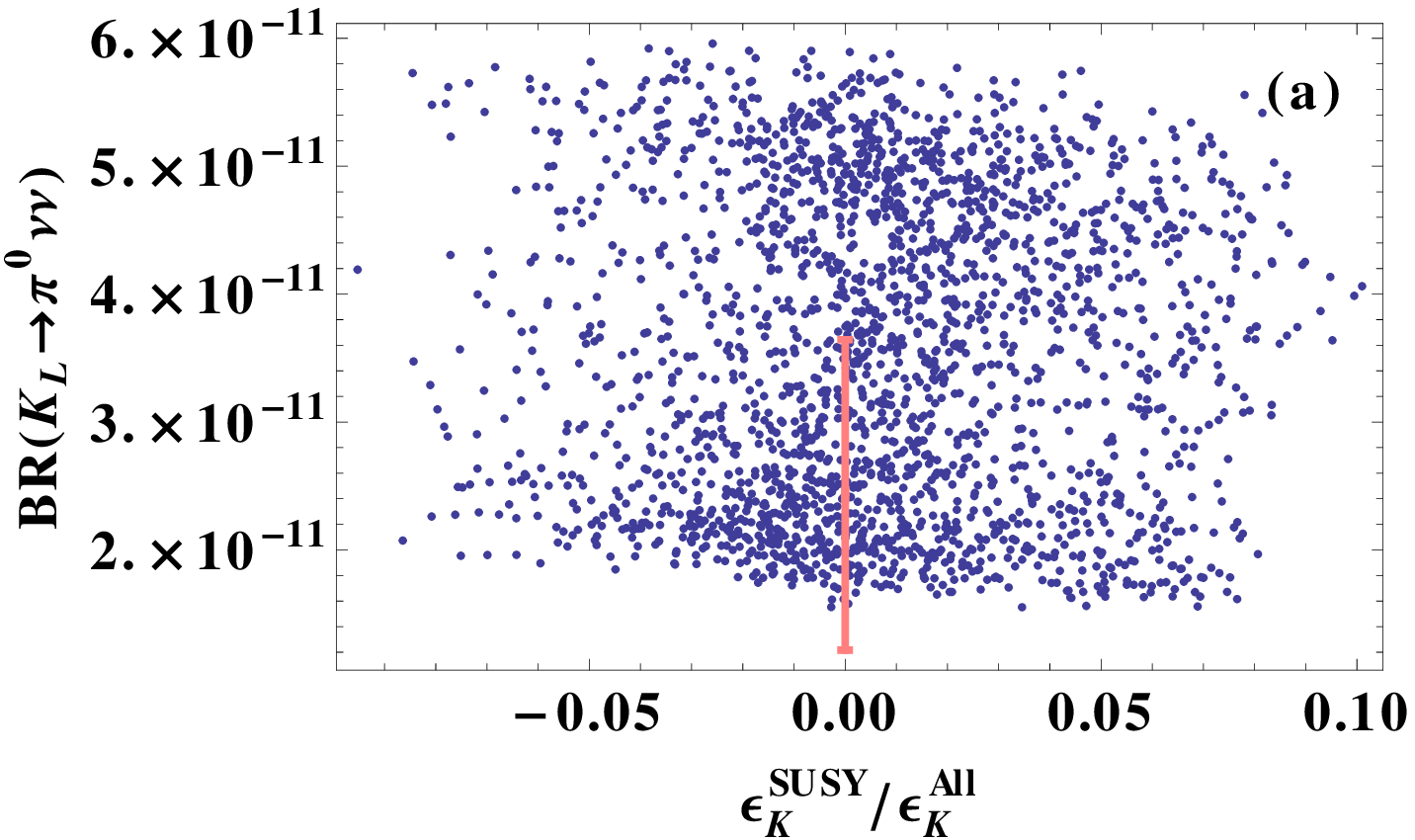}
\hspace{1cm}
\hspace{0.3cm}
\includegraphics[width=8cm]{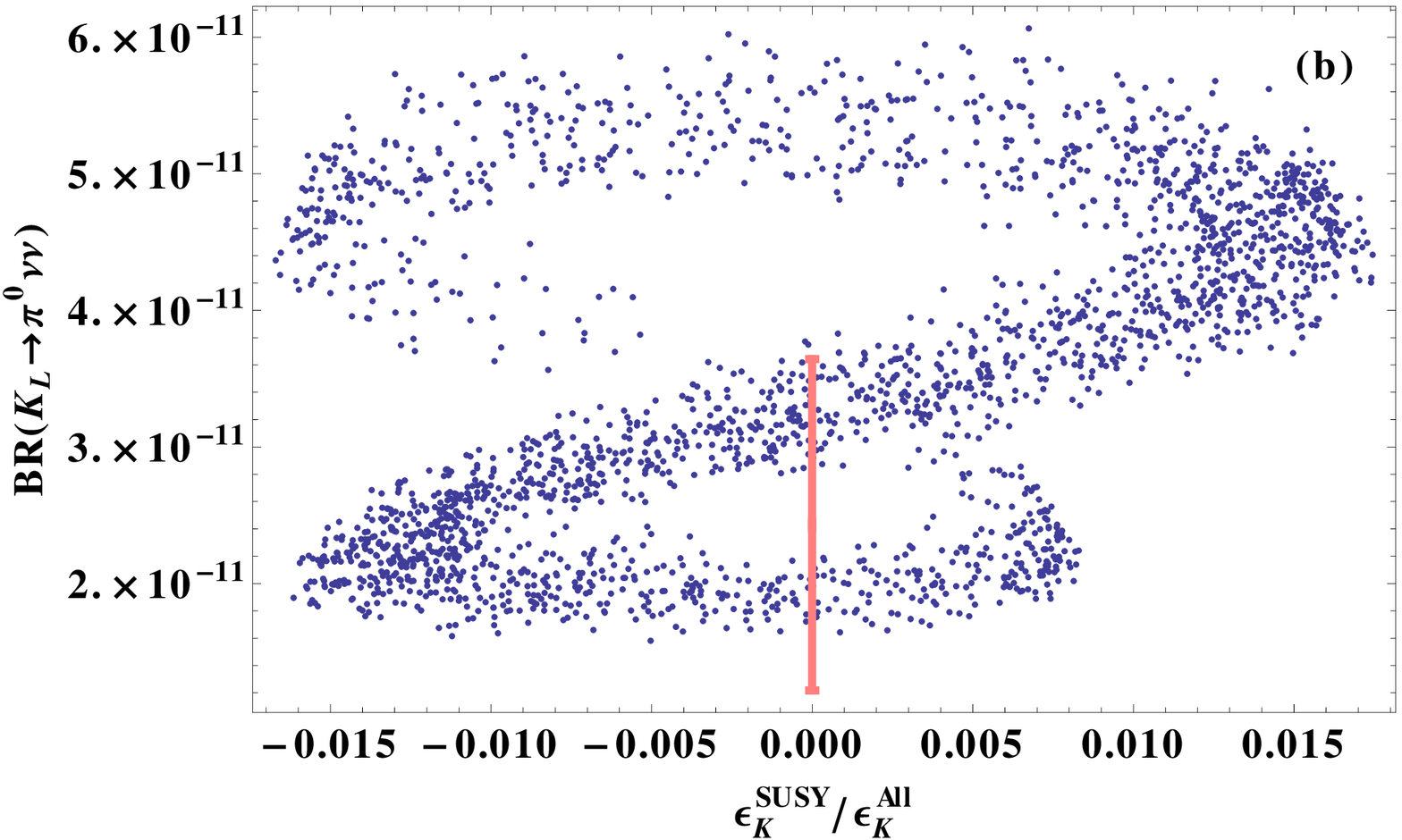}
\caption{The predicted $BR(K_L \to \pi^0 \nu  {\bar \nu})$ versus the SUSY contribution ratio 
of $\epsilon_K$  at the SUSY scale of $50$ TeV in the case of
(a) $s^u=s^d=0.3$ and (b) $s^u=0.3, \ s^d=0$.  The pink short line denotes
the SM prediction  with the error-bar of 90\%C.L. } 
\label{fig:5}
\end{figure}
%%%%%%%%%%%%%%%%%%%%%%%%%%%%%%%%%%%%%%%%%%%%%%%%%%%%%%%%%%%%
%%%%%%%%%%%%%%%%%%%%%%%%%%%%%%%%%%%%%%%%%%%%%%%%%%%%%%%%%%%
\begin{figure}[t]
\includegraphics[width=8cm]{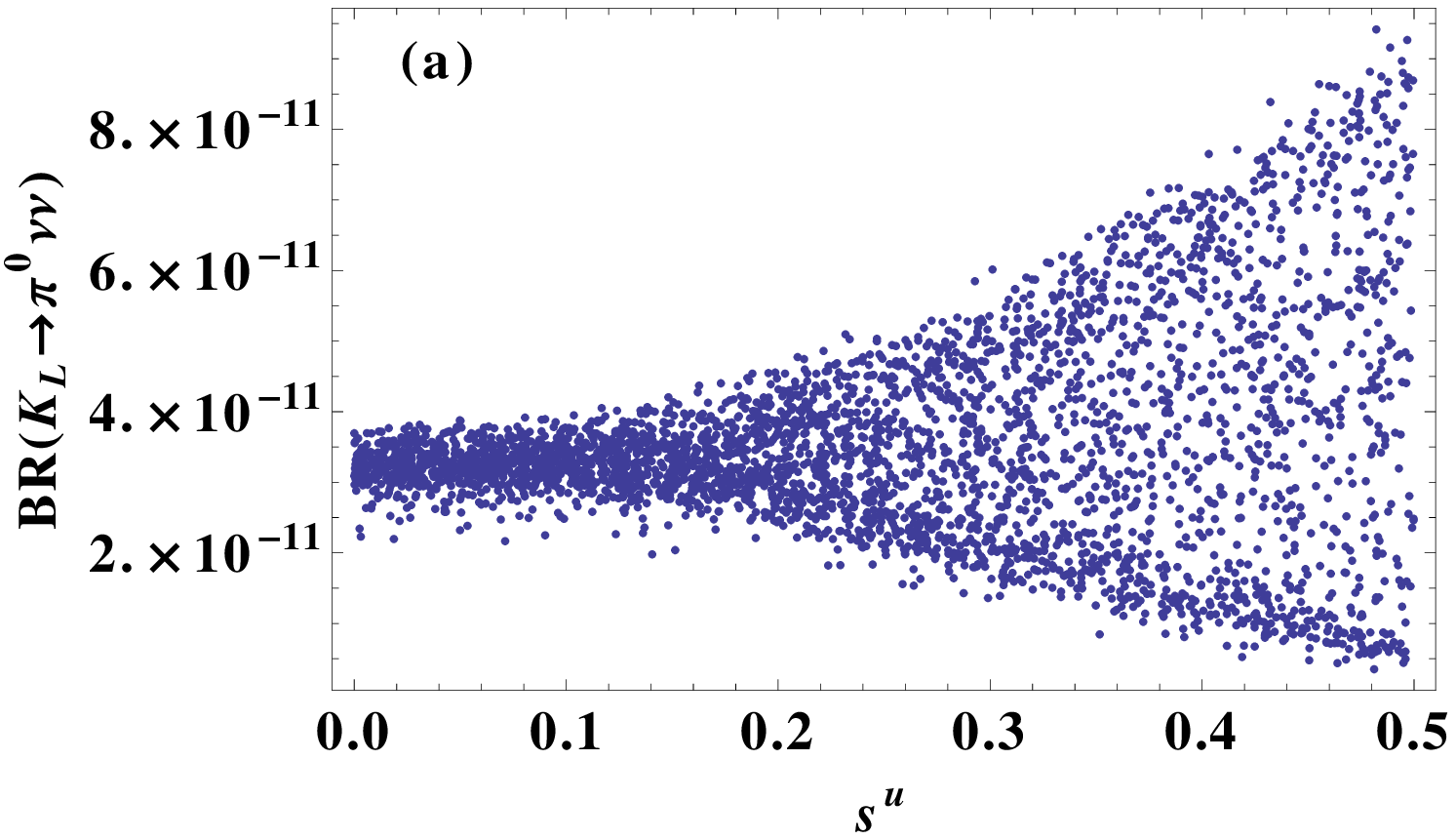}
\hspace{0.3cm}
\includegraphics[width=8cm]{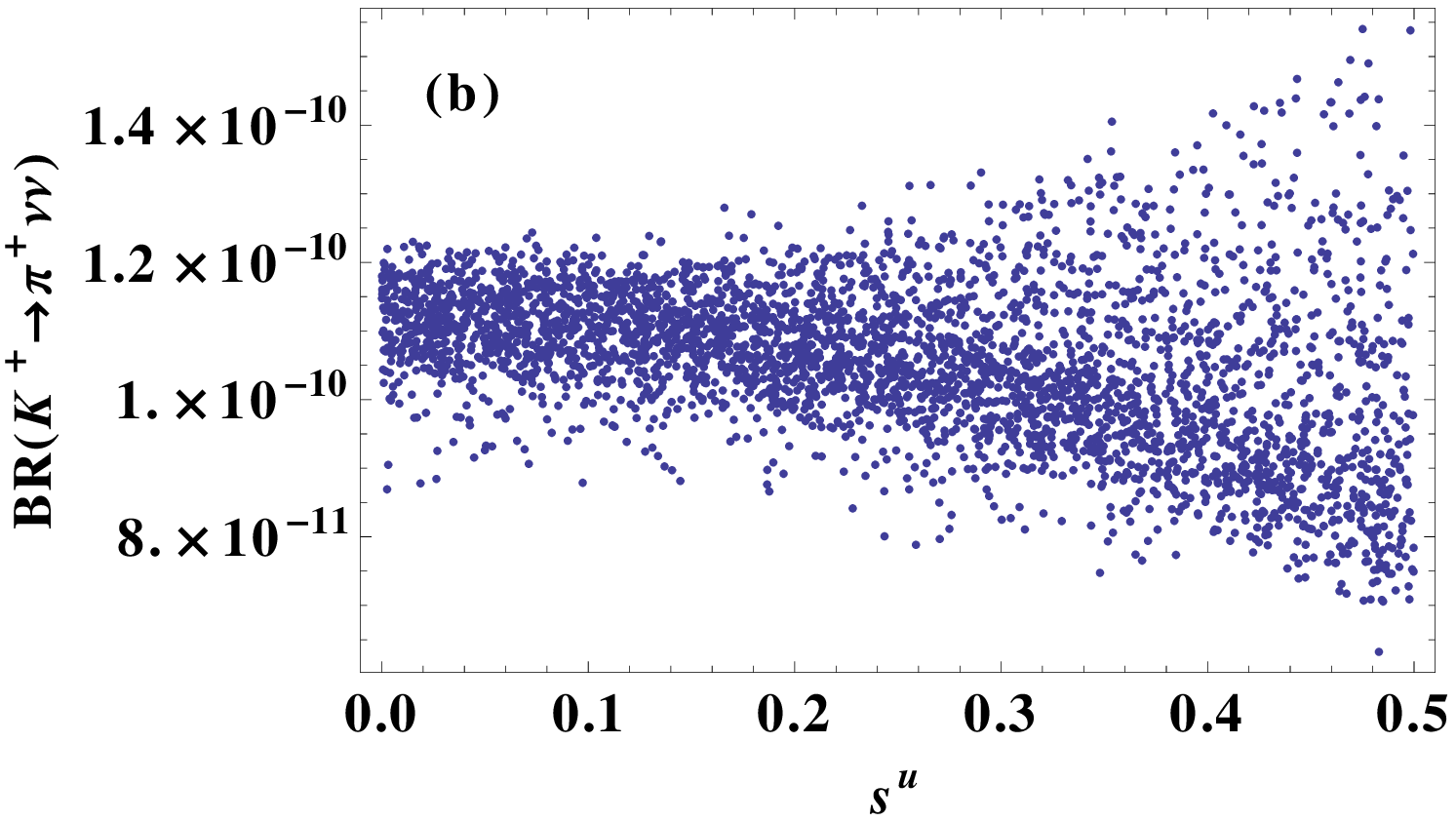}
\caption{The predicted $s^u$ dependence of (a) $BR(K_L \to \pi^0 \nu  {\bar \nu})$ and 
(b) $BR(K^+ \to \pi^+ \nu {\bar \nu})$ at  the SUSY scale of $50$ TeV, where $s^d$ is scanned in the region of $0\sim 0.3$ independent of $s^u$.} 
\label{fig:6}
\end{figure}
%%%%%%%%%%%%%%%%%%%%%%%%%%%%%%%%%%%%%%%%%%%%%%%%%%%%%%%%%%%%

%%%%%%%%%%%%%%%%%%%%%%%%%%%%%%%%%%%%%%%%%%%%%%%%%%%%%%%%%%%%%
\begin{figure}[h]
\begin{center}
\includegraphics[width=10cm]{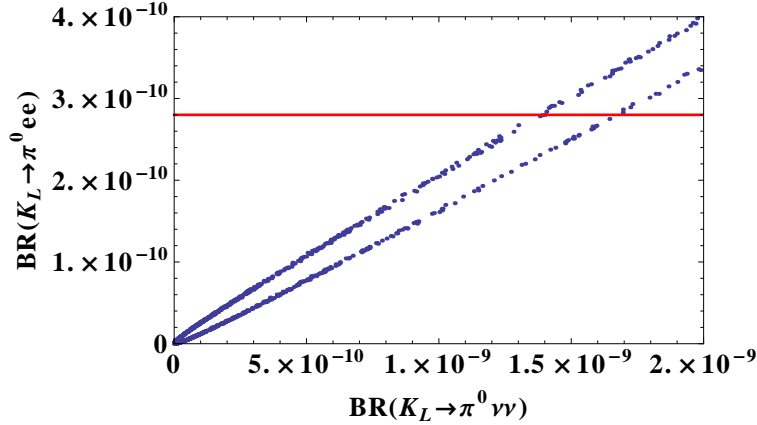}
\end{center}
\caption{The predicted $BR(K_L \to \pi^0 e^+e^-)$ versus $BR(K_L \to \pi^0 \nu  {\bar \nu})$ 
with $s^u=0\sim 0.3$ and  $s^d=0\sim 0.3$
at the SUSY scale of $10$ TeV. The red solid line denotes the upper-bound of
the branching ratio $BR(K_L \to \pi^0 e^+e^-)$.} 
\label{fig:7}
\end{figure}
%%%%%%%%%%%%%%%%%%%%%%%%%%%%%%%%%%%%%%%%%%%%%%%%%%%%%%%%%%%%
To see the  correlation between  $\epsilon_K$ and  the  predicted $K_L \to \pi^0 \nu  {\bar \nu}$ branching ratio,  we show the  branching ratio of $K_L \to \pi^0 \nu  {\bar \nu}$ 
versus the SUSY contribution of $\epsilon_K$ in Fig. \ref{fig:5}, 
in which  (a)  $s^u=s^d=0.3$ and (b) $s^u=0.3, \ s^d=0$.
We do not find any  correlation between them in the Fig. \ref{fig:5}(a), 
 where the gluino contribution to $\epsilon_K$ is still possible up to $10\%$.
 However, it is found  a correlation between them as seen in  Fig. \ref{fig:5}(b),
where  the Z penguin mediated chargino dominates both
  SUSY contributions of   $K_L \to \pi^0 \nu  {\bar \nu}$ and    $\epsilon_K$
since the down-squark mixing $s_d$ vanishes with keeping  $s_u=0.3$.
The chargino contribution  to  $\epsilon_K$ is at most $2\%$.
This correlation is understandable from the difference of the phase structure
  between the penguin diagram and the box diagram of the chargino.

%When one considers  the stop exchange diagrams, the chargino box  amplitude has   the phase factor 
%$\exp [2i(\phi^{uL(R)}_{13}-\phi^{uL(R)}_{23})]$,
%while the Z penguin mediated the chargino one has   $\exp [i(\phi^{uL(R)}_{13}-\phi^{uL(R)}_{23})]$.
%For  $\phi^{uL(R)}_{13}-\phi^{uL(R)}_{23}=\pm\pi/2$,
%the imaginary part of the   $K_L \to \pi^0 \nu  {\bar \nu}$ amplitude is maximal,
% but the imaginary part of the chargino box amplitude vanishes. 

Thus, even if  the SUSY scale is  $50$ TeV,
  $\epsilon_K$ could be  deviated  from the SM prediction in ${\cal O}(10\%)$ due to the
gluino box diagram,
whereas the chargino process deviates the branching ratio of
$K_L \to \pi^0 \nu  {\bar \nu}$ from the SM prediction in the factor two.

Fig.\ref{fig:6} shows the $s^u$ dependence of 
$BR(K_L \to \pi^0 \nu  {\bar \nu})$ 
and $BR(K^+ \to \pi^+ \nu {\bar \nu})$ 
  taking $s^u=0 \sim 0.5$
in Fig.\ref{fig:6} (a) and (b).
 We also scan  $s^d$ in the region of $0\sim 0.3$ independent of $s^u$.
In this plot, the SUSY contribution to  $\epsilon_K$ is  free ($0-40\%$), but 
the experimental  constraint of $\epsilon_K$ with the error-bar of 90\%C.L. is taken account.
The predicted $BR(K_L \to \pi^0 \nu  {\bar \nu})$ could be large up to $8\times 10^{-11}$,
and $BR(K^+ \to \pi^+ \nu {\bar \nu})$ is up to $1.5\times 10^{-10}$.
Thus, the enhancement  from  the SM prediction could be detectable 
even if  the SUSY scale is $50$TeV.

%\begin{wrapfigure}{r}{8cm}
%         		\begin{center}
%		         \includegraphics[width=8cm]{KLee-KL_10TeV_sd03su03.eps}
%		        \caption{} 
%		         \label{}
%	         \end{center}
%        \end{wrapfigure} 
%%%%%%%%%%%%%%%%%%%%%%%%%%%%%%%%%%%%%%%%%%%%%%%%%%%%%%%%
Before closing our numerical study, we would like to discuss  correlations to other quantities which are sensitive to the NP. They are $K_L \to \pi^0 e^+e^-$ process and the neutron electric dipole moment $d_n$.
The $K_L \to \pi^0 e^+e^-$ process is  induced in similar way to $K_L \to \pi^0 \nu  {\bar \nu}$.
The distinguish feature of $K_L \to \pi^0 e^+ e^-$ mode is the contribution of the photon penguin.
Moreover,  one cannot neglect the long-distance effect 
from the photon exchange process \cite{Buchalla:2003sj}. Thus, the decay amplitude of
 $K_L \to \pi^0 e^+ e^-$ has both the short-distance effect and the long-distance effect, and
 the SM prediction of the branching ratio is  around $ 3\times10^{-11}$, which is comparable to the
 SM prediction of   $K_L \to \pi^0 \nu  {\bar \nu}$.
Since our interest here is to check whether the SUSY effect does not exceed the experimental bound of $K_L \to \pi^0 e^+ e^-$, we only consider the short distance contribution in our analysis.
The experimental bound of the branching ratio $K_L \to \pi^0 e^+ e^-$ is 
${\rm BR}(K_L \to \pi^0 e^+ e^-)_{\rm exp}< 2.8 \times 10^{-10}$ \cite{PDG}.
In the Fig.\ref{fig:7}, the predicted $BR(K_L \to \pi^0 e^+e^-)$ vs. 
$BR(K_L \to \pi^0 \nu  {\bar \nu})$ plane are plotted
with $s^u=0\sim 0.3$ and  $s^d=0\sim 0.3$ at the $10$TeV scale of the SUSY. 
There are  two predicted lines in this figure. 
Because the decay amplitude $A(K_L \to \pi^0 e^+ e^- )$ is described by  the sum of the SM and the SUSY contributions, there are two ways of taking the relative phase of $\pm$
 such as $A(K_L \to \pi^0 e^+ e^- )=A(K_L \to \pi^0 e^+ e^- : {\rm SM})
\pm A(K_L \to \pi^0 e^+ e^- : {\rm SUSY})$,  
which has two solutions giving the same absolute value  of the decay amplitude.
Then,  we have two predicted values of $BR(K_L \to \pi^0 e^+e^-)$
for the certain  $BR(K_L \to \pi^0 \nu  {\bar \nu})$.
%%%%%%%%%%%%%%%%%%%%%%%%%%%%%%%%%%%%%%%%%
 The both decay processes are dominated by  the $Z$ penguin mediated charginos,
 then, the branching ratios are determined  by the final state  couplings of 
$Z\nu\bar\nu$ and $Ze^+e^-$, that is,  the weak charges $Q_{ZL}^{(\nu)}$ and   $Q_{ZL}^{(e)}$.
Moreover,  three flavors of neutrinos  are summed for  $K_L \to \pi^0 \nu  {\bar \nu}$.
Therefore, $BR(K_L \to \pi^0 \nu  {\bar \nu})$ is significantly larger 
than $BR(K_L \to \pi^0 e^+e^-)$.
 On the other hand, in the SM, there are some contributions to $K_L \to \pi^0 e^+e^-$ 
such as the photon exchange processes.
So, $BR(K_L \to \pi^0 e^+e^-)$ is comparable to $BR(K_L \to \pi^0 \nu  {\bar \nu})$
in the SM.
%%%%%%%%%%%%%%%%%%%%%%%%%%%%%%%%%%%%%%%%%
In conclusion, the experimental upper bound of $BR(K_L \to \pi^0 e^+e^-)$ 
  excludes the region larger than  $BR(K_L \to \pi^0 \nu  {\bar \nu})=1.7\times 10^{-9}$.
  However, if the long-distance effect is properly included \cite{Buchalla:2003sj}, 
this constraint becomes  somewhat tight or loose depending on the relative sign between
 the SUSY contribution and the long-distance one.
 
%%%%%%%%%%%%%%%%%%%%%%%%%%%%%%%%%%%%%%%%%%%%%%%%%%%%%%%%%%%%%%%%%%%%%%%%%%%%%%

The neutron electric dipole moment (EDM) $d_n$ is well known 
as  the sensitive probe for the NP, and so we have studied the correlation between the neutron EDM  and the $K \to \pi^0 \nu  {\bar \nu}$ branching ratio. 
%The experimental upper bound of $d_n$ are given  $d_n < 0.29 \times 10^{-25}{\rm e \cdot cm}$.
%though the predicted neutron EDM can be exceed this upper bound around 3 times, 
It is found that our predicted $K \to \pi^0 \nu  {\bar \nu}$ does not correlate with $d_n$.
%and is not constrained by the $d_n$ ,
Suppose the SUSY contribution to the chromo-EDM of quarks through 
the gluon penguin mediated gluino
 \cite{Hisano:2003iw}-\cite{Fuyuto:2013gla}, where 
 the left-right mixing term of the down-squark is dominant.
 In our SUSY mass spectra, the  left-right mixing is suppressed as discussed in section 3.
 Moreover, 
 the CP violating phase dependence of $d_n$ comes from the  down-squark mixing matrix
whereas  the phase of $K \to \pi^0 \nu  {\bar \nu}$ comes from the up-squark mixing matrix.
 Namely, those  phase dependences  are  completely different each other.
Therefore, we do not take account of the constraint from 
the experimental upper bound of the neutron EDM  in our analyses.

%%%%%%%%%%%%%%%%%%%%%%%%%%%%%%%%%%%%%%%%%%%%%%%%%%%%%%%%%%%
%%%%%%%%%%%%%%%%%%%%%%%%%%%%%%%%%%%%%%%%%%%%%%%%%%%%%%%%%%%
\section{Summary}
We have studied the contribution of the high-scale SUSY to the 
 $K_L \to \pi^0 \nu{\bar \nu}$ and $K^+ \to \pi^+ \nu{\bar \nu}$ processes
 by correlating with the CP violating parameter $\epsilon_K$.
These rare decays  have important role of the decision of the CP phase in the  CKM matrix, 
furthermore, they are also sensitive to the flavor structure of the NP. 

Taking account of the recent LHC results for the Higgs discovery and the SUSY searches,
 we consider the hight-scale SUSY at the $10-50$TeV scale.
Then,  we have discussed the  SUSY effects  to
  $K^+ \to \pi^+ \nu {\bar \nu}$, $K_L \to \pi^0 \nu  {\bar \nu}$ and $\epsilon_K$
  in the framework of the mass eigenstate basis of the SUSY particles assuming the non-minimal squark (slepton) flavor mixing.

We have calculated  the SUSY contribution to the branching ratios 
of  $K_L \to \pi^0 \nu  {\bar \nu}$ and $K^+ \to \pi^+ \nu {\bar \nu}$, where phase parameters are constrained by the observed $\epsilon_K$. 
The Z penguin mediated chargino  dominates the SUSY contribution
 for these decays.
 At the $10$ TeV scale of the SUSY, its contribution can enhance the  branching ratio
of  $K_L \to \pi^0 \nu {\bar \nu}$
in eight times compared with the SM predictions whereas
  the predicted branching ratio $BR(K^+ \to \pi^+ \nu {\bar \nu})$
increases up to three times of the SM prediction
in the case of the up-squark mixing $s^u=0.1 $.

We have investigated the  correlation between  $\epsilon_K$ and  
the  $K_L \to \pi^0 \nu  {\bar \nu}$ branching ratio.
Since the gluino box diagram
dominates the SUSY contribution of   $\epsilon_K$ up to $30\%$,
there is no correlation between them.
However,
if the down-squark mixing is neglected compared with  the up-squark mixing,
the chargino process dominates both
  SUSY contributions of   $K_L \to \pi^0 \nu  {\bar \nu}$ and    $\epsilon_K$.
Then, it is found  a correlation between them, but
 the chargino contribution  to  $\epsilon_K$ is at most $3\%$.
It is concluded that  $\epsilon_K$ could be  deviated significantly from the SM prediction in ${\cal O}(10\%)$ due to the gluino box process,
whereas the chargino process could enhance the branching ratio of
$K_L \to \pi^0 \nu  {\bar \nu}$  in  several times from the SM prediction.

Our predicted  branching ratios depend on  the  mixing angle $s^u$ 
significantly.
 The observed upper bound of $BR(K^+ \to \pi^+ \nu {\bar \nu})$
 gives the  constraint for the predicted $BR(K_L \to \pi^0 \nu  {\bar \nu})$
at $s^u$ larger than $0.2$.

 Even if the SUSY scale is  $50$ TeV, the chargino process still enhances  the branching ratio of
$K_L \to \pi^0 \nu  {\bar \nu}$ from the SM prediction in the factor two, and 
 the $\epsilon_K$ is deviated  from the SM prediction in ${\cal O}(10\%)$
 unless  the down-squark mixing $s^d$ is suppressed.
 
We also discuss correlations to the $K_L \to \pi^0 e^+e^-$ process and the neutron electric dipole moment $d_n$ which are sensitive to the NP. 

We expect  the measurement of these processes will be improved by the J-PARC KOTO experiment and CERN NA62 experiment in the near future.

%%%%%%%%%%%%%%%%%%%%%%%%%%%%%%%%%%%%%%%%%%%%%%%%%%%%%%%%%%%

%%%%% acknowledgement %%%%%
\vspace{0.5 cm}
\noindent
{\bf Acknowledgment}

\vskip 0.3 cm
We would like to thank T. Kurimoto for the  comment at the early stage of this work.
This work is supported by JSPS Grants-in-Aid for Scientific Research,
 24654062 and 25-5222.

%%%%%%%%%%%%%%%%%%%%%%%%%%%%%%%%%%%%%%%%%%%%%%%%%%%%%%%%%%%%
%%%%%%%%%%%%%%%%%%%%%%%  Appendices %%%%%%%%%%%%%%%%%%%%%%%%
%%%%%%%%%%%%%%%%%%%%%%%%%%%%%%%%%%%%%%%%%%%%%%%%%%%%%%%%%%%%
\newpage
\appendix{}
\section*{Appendix A : SUSY Spectrum}

In the framework of the MSSM,
one  obtains the SUSY particle spectrum which is consistent with the observed Higgs mass.
The numerical analyses have been given in Refs.
\cite{Delgado:2013gza, Giudice:2006sn}.
At the SUSY breaking scale $\Lambda$, the quadratic terms in the MSSM potential is given as
\begin{equation}
 V_2=m_1^2 |H_1|^2+ m_2^2 |H_2|^2+m_3^2 (H_1\cdot H_2+ h.c.) \ .
\end{equation}
The mass eigenvalues at  the $H_1$ and $\tilde H_2\equiv \epsilon H_2^*$ system are given
\begin{equation}
 m_\mp^2 = \frac{m_1^2+m_2^2}{2}\mp \sqrt{\left (\frac{m_1^2-m_2^2}{2}\right )^2+m_3^4} \ .
\end{equation}
Suppose that the MSSM matches with the SM at the SUSY mass scale $Q_0\equiv m_0$.
Then, the smaller one $m_{-}^2$ is identified to be the mass squared of the SM Higgs $H$ with the tachyonic mass.
The larger one $m_{+}^2$ is  the mass squared of the orthogonal combination ${\cal H}$, which
is decoupled from the SM at $Q_0$, that is, $m_{{\cal H}}\simeq Q_0$
. Therefore, we have
\begin{eqnarray}
 m_{-}^2=-m^2(Q_0)\ , \qquad  m_{+}^2=m_{\cal H}^2(Q_0) =m_1^2+m_2^2+m^2 \ ,
\end{eqnarray}
with 
\begin{eqnarray}
 m_{3}^4=(m_1^2+m^2)(m_2^2+m^2) \ ,
\end{eqnarray}
which leads to the mixing angle between $H_1$ and $\tilde H_2$, $\beta$ as follows:
\begin{eqnarray}
 \tan^2\beta=\frac{m_1^2+m^2}{m_2^2+m^2} \ , \qquad 
 H=  \cos\beta H_1 +\sin\beta \tilde H_2  \ , \qquad
 {\cal H} =-\sin\beta H_1 +\cos\beta \tilde H_2  \ .
\end{eqnarray}
Thus,  the Higgs mass parameter $m^2$ is expressed in terms of $m_1^2$, $m_2^2$ and $\tan\beta$:
\begin{eqnarray}
m^2=\frac{m_1^2-m_2^2\tan^2\beta}{\tan^2\beta -1} \ .
\end{eqnarray}
Below the  $Q_0$ scale, in which the SM emerges, the scalar potential is  the SM one as follows:
\begin{equation}
 V_{SM}=-m^2 |H|^2+\frac{\lambda}{2} |H|^4 \ .
\end{equation}
Here, the Higgs coupling $\lambda$ is given in terms of the SUSY parameters
at the leading order as
 \begin{equation}
 \lambda(Q_0)=\frac{1}{4} (g^2+g'^2)\cos^2 2\beta +\frac{3h_t^2}{8\pi^2} X_t^2 \left (1- \frac{X_t^2}{12}\right ) \ , \qquad
 X_t=\frac{A_t(Q_0)-\mu(Q_0)\cot\beta}{ Q_0 } \ ,
\end{equation}
and $h_t$ is the top Yukawa coupling of the SM.
 The parameters $m_2$ and $\lambda$ run with the SM Renormalization Group Equation down to the electroweak scale $Q_{EW}=m_H$, and then give
\begin{equation}
 m_H^2=2m^2(m_H)=\lambda(m_H)v^2\ .
\end{equation}
It is easily seen  that the VEV of Higgs, $\langle H \rangle $ is $v$,  and $\langle {\cal H}\rangle=0$,
taking account of $\langle H_1\rangle =v\cos\beta$ and $\langle H_2\rangle =v\sin\beta$,
 where $v=246$GeV.

Let us  fix $m_H=126$GeV, which gives $\lambda(Q_0)$ and $m^2(Q_0)$. This experimental input constrains the SUSY mass spectrum of the MSSM.
We consider the some universal soft breaking parameters at the SUSY breaking scale $\Lambda$ as follows:
\begin{eqnarray}
&&m_{\tilde Q_i}(\Lambda)=m_{\tilde U^c_i}(\Lambda)=m_{\tilde D^c_i}(\Lambda)=
m_{\tilde L_i}(\Lambda)=m_{\tilde E^c_i}(\Lambda)=m_0^2 \ (i=1,2,3) \ , \nonumber \\
&&M_1(\Lambda)=M_2(\Lambda)=M_3(\Lambda)=m_{1/2}  \ , \qquad 
m_1^2({\Lambda})= m_2^2({\Lambda})=m_0^2 \ , \nonumber \\
&&A_U({\Lambda})=A_0 y_U(\Lambda)\ , \quad A_D({\Lambda})=A_0 y_D(\Lambda)\ , 
\quad A_E({\Lambda})=A_0 y_E(\Lambda)\ .
\end{eqnarray}
Therefore, there is no flavor mixing at $\Lambda$ in the MSSM.  
However, in order to consider the non-minimal flavor mixing framework, we allow the off diagonal components of the squark mass matrices at the $10\%$ level, which leads to the
flavor mixing of order $0.1$.
 We take these flavor mixing angles as free parameters  at low energies.

Now, we have the SUSY five parameters, $\Lambda$, $\tan\beta$, $m_0$, $m_{1/2}$, $A_0$,
where $Q_0=m_0$. In addition to these parameters, we take $\mu=Q_0$.
By fixing $\Lambda$, $Q_0$ and $\tan\beta$, we tune
$m_{1/2}$ and  $A_0$ in order to obtain $m^2(Q_0)$ and $\lambda_H(Q_0)$ which realize the correct electroweak vacuum with  $m_H=126$GeV.
Then, we  obtain the SUSY particle spectrum.
We consider the two case of $Q_0=10$ TeV and $50$ TeV.
The input parameter set and the obtained SUSY mass spectra at $Q_0$ are  summarized in Table 1,
where we use $\overline m_t (m_t)=163.5\pm 2$ GeV \cite{PDG,UTfit}.
% These parameter sets are easily found  from the  work  in Ref.\cite{Delgado:2013gza}.
\vskip 0.5 cm
\begin{table}[h!]
\begin{center}
\begin{tabular}{|l|l|}\hline
 \quad Input at $\Lambda$ and $Q_0$ & \qquad  \qquad \qquad \quad Output at $Q_0$ \\
\hline
%   & \\
 at $\Lambda=10^{17}$ GeV,  & $m_{\tilde g}=12.8$ TeV,  \ $m_{\tilde W}=5.2$ TeV,  
\ $m_{\tilde B}=2.9$ TeV\\ 
 \quad $m_0=10$ TeV,  & $m_{\tilde b_L}=m_{\tilde t_L}=12.2$ TeV  \\
 \quad $m_{1/2}=6.2$ TeV, &  $m_{\tilde b_R}=14.1$ TeV, \  $m_{\tilde t_R}=8.4$ TeV\\
 \quad  $A_0=25.803$ TeV; & $m_{\tilde s_L , \tilde d_L}=m_{\tilde c_L , \tilde u_L}=15.1$ TeV \\
 at $Q_0=10$ TeV, &   $m_{\tilde s_R,\tilde d_R}\simeq m_{\tilde c_R,\tilde u_R}=14.6$ TeV,  
\ $m_{{\cal H}}=13.7$ TeV  \\
\quad  $\mu=10$ TeV,&  $m_{\tilde \tau_{L}}=m_{\tilde {\nu_\tau}_L}=10.4$ TeV, \  
$m_{\tilde \tau_R}=9.3$ TeV 
\\
 \quad $\tan\beta=10$&  
 $m_{\tilde \mu_L, \tilde e_{L}}=m_{\tilde {\nu_\mu}_L, \tilde {\nu_e}_L}=10.8$ TeV, 
 \ $m_{\tilde \mu_R, \tilde e_R}=10.3$ TeV \\
  &  $X_t=-0.22$, \quad  $\lambda_H=0.126$ \\
 &  $m_1^2=1.84857\times 10^8{\rm GeV}^2$, \ \  $m_2^2=1.83996\times 10^6 {\rm GeV}^2$,
 \ \   $m^2=8691 {\rm GeV}^2$ 
\\
\hline
%  & \\
at $\Lambda=10^{16}$ GeV,  & $m_{\tilde g}=115.6$ TeV,  \ $m_{\tilde W}=55.4$ TeV,  \ $m_{\tilde B}=33.45$ TeV\\ 
\quad  $m_0=50$ TeV,  & $m_{\tilde b_L}=m_{\tilde t_L}=100.9$ TeV  \\
 \quad $m_{1/2}=63.5$ TeV, &  $m_{\tilde b_R}=104.0$ TeV, \  $m_{\tilde t_R}=83.2$ TeV\\
 \quad $A_0=109.993$ TeV; & $m_{\tilde s_L , \tilde d_L}=m_{\tilde c_L , \tilde u_L}=110.7$ TeV, \    $m_{\tilde s_R,\tilde d_R}=110.7$ TeV\\
at $Q_0=50$ TeV,   &    $m_{\tilde c_R,\tilde u_R}=105.0$ TeV,  \ $m_{{\cal H}}=83.1$ TeV  \\
\quad  $\mu=50$ TeV,&  $m_{\tilde \tau_{L}}=m_{\tilde {\nu_\tau}_L}=63.6$ TeV, \  
$m_{\tilde \tau_R}=54.6$ TeV 
\\
 \quad $\tan\beta=4$&  
 $m_{\tilde \mu_L, \tilde e_{L}}=m_{\tilde {\nu_\mu}_L, \tilde {\nu_e}_L}=63.8$ TeV, 
 \ $m_{\tilde \mu_R, \tilde e_R}=55.0$ TeV \\
  &  $X_t=-0.65$, \quad  $\lambda_H=0.1007$ 
\\
 &  $m_1^2=6.49990\times 10^9{\rm GeV}^2$, \ \  $m_2^2=4.06235\times 10^8 {\rm GeV}^2$,
 \ \   $m^2=8840 {\rm GeV}^2$ 
\\
\hline
\end{tabular}
\end{center}
\caption{Input parameters at $\Lambda$ and  the obtained SUSY spectra at
$Q_0=10$ and $50$TeV.}
\label{tab:inputparameters}
\end{table}

\section*{Appendix B : Z penguin amplitude mediated charginos}

We present the  expression for the Z  penguin amplitude mediated the chargino, 
$P_{\rm ZL}^{sd}(\chi^{\pm})$  in our basis \cite{GotoNote} as follows:

%%%%%%%%%%%%%%%

\begin{align}
P_{\rm ZL}^{sd}(\chi^{\pm})
&=\frac{g_2^2}{4m_W^2}
\sum_{\alpha,\beta.I,J}
(\Gamma _{CL}^{(d)\dagger})^I_{\alpha d}
(\Gamma _{CL}^{(d)})_J^{\beta s}
\Big\{\delta^J_I (U_+^\dagger)_\beta^1 (U_+)^\alpha_1  
\ [\log x^{\mu_0}_I + f_2(x^I_\alpha, x^I_\beta) ]\\
&-2 \delta^J_I (U_-^\dagger)_\beta^1 (U_-)^\alpha_1 \sqrt{x^I_\alpha x^I_\beta}f_1(x^I_\alpha, x^I_\beta)
-\delta^{\alpha}_{\beta}\left( \tilde{\Gamma}_L^{(u)}\right)^J_I f_2(x_I^{\alpha},x_J^{\alpha}) \Big\} \ ,
\end{align}  
where
\begin{equation}
(\Gamma _{CL}^{(d)})_I^{\alpha q}\equiv 
(\Gamma _{L}^{(u)}  V_{\rm CKM})_I^q (U_+)^\alpha_1
+\frac{1}{g_2}(\Gamma _{R}^{(u)} \hat f_U V_{\rm CKM})_I^q (U_+)^\alpha_2 \ ,
\end{equation}  
and
\begin{equation}
\left(\tilde{\Gamma}_L^{(u)} \right)_I^{ \ J}
\equiv
\left( \Gamma _{L}^{(u)} \Gamma _{L}^{(u)\dagger}
\right)_I^{ \ J} \ ,
\end{equation}  
with
 $q=s, d$,  $I=1-6$ for up-squarks,  and $\alpha=1,2$  for charginos.
 The  $V_{\rm CKM}$ is the CKM matrix, 
and   $U_\pm$ are the  $2\times 2$ unitary matrices which diagonalize 
  the  chargino mass matrix  $M_C$:
\begin{align}
U_-^{\dagger}M_C U_+
=
-{\rm diag}M_C^{\alpha}\ , \ \ \ \ \ \ (\alpha = 1,2) \ .
\end{align}
The $\hat f_U$ denotes the yukawa coupling defined by $\hat f_U v \sin{\beta}=
{\rm diag}(m_u,m_c,m_t)$.
The loop integral functions are given as:  
\begin{align}
f_n(x,y)
=\frac{1}{x-y}\left( \frac{x^n {\rm log}x}{x-1}-\frac{y^n {\rm log}y}{y-1} \right) \ ,
\end{align}  
with
\begin{align}
x^I_\alpha=\frac{m_{\chi_{\alpha}}^2}{\tilde m_I^2} \ , \qquad  
x^{\mu_0}_I=\frac{\tilde m_I^2}{\mu_0^2}
 \ ,
\end{align}
where  $\mu_0=Q_0$  is taken in our framework.  

\newpage
%%%%%%%%%%%%%%%%%%%%%%%%%%%%%%%%%%%%%%%%%%%%%%%
%%%%%%%%  Regerences %%%%%%%%%%%%%%%%%%%%%%%%%%
%%%%%%%%%%%%%%%%%%%%%%%%%%%%%%%%%%%%%%%%%%%%%%%

\end{document}